\documentclass[12pt]{article}
\pdfoutput=1
\usepackage{amsmath,amssymb,bm,graphicx,cite,float,booktabs,tabularx,array,microtype}
\usepackage[margin=25mm]{geometry}
\usepackage[colorlinks=true,allcolors=blue]{hyperref}
\title{Real-Frequency Scattering and Superradiant Response across a Kerr--Bertotti--Robinson Bridge}
\hypersetup{pdftitle={Real-Frequency Scattering and Superradiant Response across a Kerr--Bertotti--Robinson Bridge},pdfauthor={Hai Huang, Xudong Sun and Juhua Chen}}
\numberwithin{equation}{section}

\newcommand{\ii}{\mathrm{i}}
\newcommand{\dd}{\mathrm{d}}
\newcommand{\noo}{\nu_{\mathrm{o}}}
\newcommand{\nii}{\nu_{\mathrm{i}}}
\newcommand{\rro}{r_{\mathrm{o}}}
\newcommand{\rri}{r_{\mathrm{i}}}
\newcommand{\qb}{q_{\mathcal B}}
\newcolumntype{Y}{>{\raggedright\arraybackslash}X}
\makeatletter\newcommand{\papertitle}{\@title}\makeatother
\begin{document}
\hypersetup{pageanchor=false}
\pagenumbering{roman}
\begin{titlepage}
\thispagestyle{empty}
\begin{center}
\vspace*{8mm}
{\Large\bfseries\papertitle}\par\vspace{9mm}
Hai Huang$^{\dagger,1}$, Xudong Sun$^{2}$ and Juhua Chen$^{\ast,3}$\par\vspace{5mm}
{\small\itshape $^1$College of Mechanical Engineering, Guiyang University,\\
103 Jianlong Rd., Guiyang, Guizhou 550005, China}\par\vspace{2mm}
{\small\itshape $^2$Department of Theoretical Physics, Kunming University of Science and Technology,\\
727 Jingming South Rd., Kunming, Yunnan 650091, China}\par\vspace{2mm}
{\small\itshape $^3$Department of Physics, Hunan Normal University,\\
36 Lushan Rd., Changsha, Hunan 410081, China}
\end{center}
\vspace{6mm}\hrule\vspace{3mm}
\noindent\textbf{Abstract}\par
We study the scattering and superradiant amplification of a neutral, minimally coupled massless scalar field in Kerr–Bertotti–Robinson spacetime. The
region considered extends outward from the black-hole outer horizon, across the coordinate bridge, to the inner horizon of a neighbouring spacetime
region. We calculate reflection and transmission across this entire region and use flux conservation to determine when the reflected wave is amplified.
Imposing transparent boundary conditions at the coordinate endpoint \(r\to+\infty\) gives a magnetic-field cutoff for superradiant amplification within
that boundary model. We find that retaining the neighbouring region beyond this endpoint allows amplification to persist beyond this cutoff. Waves can
still pass through the bridge when a local barrier is present. For non-extremal horizons, the amplification falls linearly to zero at either edge of the
amplification interval, while the peaks found numerically are associated with damped resonances. These results show that transmission and superradiant
amplification depend on the full region retained and its endpoint conditions.

\par\noindent\textbf{Keywords}: Kerr--Bertotti--Robinson geometry; scalar scattering; superradiance; horizon flux.
\vspace{3mm}\hrule\vfill
\noindent{\small $^\dagger$\href{mailto:haihuang@gyu.edu.cn}{haihuang@gyu.edu.cn}\quad
$^\ast$\href{mailto:jhchen@hunnu.edu.cn}{jhchen@hunnu.edu.cn}}
\end{titlepage}
\hypersetup{pageanchor=true}
{\small\setlength{\parskip}{0pt}\hypersetup{linkcolor=black}\tableofcontents}
\clearpage\pagenumbering{arabic}

\section{Introduction}\label{sec:introduction}

Wave scattering by a rotating black hole depends not only on the geometry near the horizon, but also on the region through which the wave propagates and
the boundary conditions imposed at its ends. This is particularly important in Kerr--Bertotti--Robinson (Kerr--BR) spacetime. Unlike the usual
asymptotically flat Kerr geometry, the coordinate limit $r\to+\infty$ in Kerr--BR is a regular bridge that can be continued into a neighbouring spacetime
region. A wave that reaches this bridge can therefore propagate further, rather than being required to satisfy a boundary condition there. This raises a
natural question: how does retaining the neighbouring region change the reflection, transmission and superradiant amplification of scalar waves?

The Bertotti--Robinson solution describes an electromagnetic universe with a geometry different from asymptotically flat
spacetime~\cite{Bertotti:1959uniform,Robinson:1959maxwell}. Its rotating Kerr--BR deformation is an exact solution of the Einstein--Maxwell
equations~\cite{Podolsky:2025tle}. Recent work has further clarified its relation to the wider Kerr--Newman--BR family, including the distinction between
this solution and the uncharged rotating member of that family
\cite{Ovcharenko:2026genuine}. The geometry also possesses
separation properties that make it possible to study wave propagation through ordinary differential equations. These properties are related to the hidden
symmetries of rotating spacetimes~\cite{Carter:1968separable,Teukolsky:1973perturbations, FrolovKrtousKubiznak:2017hidden}. In particular, for a neutral
massless scalar field, the tracelessness of the Maxwell stress tensor gives $R[g]=0$. A conformal field rescaling then leads to exactly separated angular
and radial equations
\cite{Gray:2025lwy}.

Superradiance is a characteristic wave phenomenon in rotating black-hole spacetimes. When a wave interacts with a rotating horizon, the reflected wave
can carry more flux than the incident wave. In asymptotically flat Kerr spacetime, this process is described by scattering between horizon modes and
modes at spatial infinity. Its frequency range is determined by the wave frequency measured relative to the rotating horizon
\cite{Starobinskii:1974amplification,Unruh:1974secondquantization,
TeukolskyPress:1974interaction,FuttermanHandlerMatzner:1988scattering, Glampedakis:2001cx}. The relation between wave amplitudes and physical fluxes is
essential in this analysis
\cite{Unruh:1976absorption,Page:1976emission,Macedo:2013afa}.
It is also important to distinguish superradiant amplification from superradiant instability: amplification describes the response to an incident wave,
whereas an instability requires additional conditions that allow the amplified wave to return and interact with the black hole
repeatedly~\cite{Brito:2015oca}.

The scattering problem in Kerr--BR has a different structure because the coordinate bridge replaces the usual asymptotic infinity. The choice of what
lies beyond the bridge therefore becomes part of the wave problem. A transparent boundary condition at $r\to+\infty$ describes one possible model, in
which the neighbouring region is replaced by an auxiliary wave channel. Retaining the neighbouring region up to its inner horizon gives a different
problem, in which the wave propagates through the extended geometry before reaching the second endpoint. The two descriptions need not have the same
transmission or amplification range. This distinction is also relevant when comparing different magnetized black-hole geometries. A test magnetic field
on Kerr, the magnetized Ernst--Wild solutions and Kerr--BR are distinct physical
constructions~\cite{Wald:1974np,Ernst:1976mzr, Ernst:1976bsr,Aliev:1989wx}. Magnetic confinement can support superradiant instabilities in suitable
settings
\cite{Konoplya:2008magnetic,Brito:2014magnetic}, while outer
ergoregions can affect the choice of stationary time generator
\cite{Gibbons:2013yq}. Mirror and anti-de Sitter examples further
show how the boundary conditions influence the subsequent evolution of amplified waves~\cite{Cardoso:2004nk,Cardoso:2004hs}.

Several aspects of Kerr--BR spacetime have been studied in recent years, including geodesics, shadows and photon rings
\cite{Wang:2025kbrshadow,Wan:2026kbrphotonrings}, as well as
approximate ringdown and charged massive scalar clouds
\cite{Mustafa:2026gly,Xu:2026ags}. Other constructions involving
electromagnetic universes and rotating geometries describe related but different solutions~\cite{Ma:2026ima,Ma:2026otg}. The global extension across
$r\to\infty$ has also received considerable attention. Ovcharenko established the extension using the reciprocal coordinate $y=1/r$ for the wider
Kerr--Newman--BR family~\cite{Ovcharenko:2026global}. Zhou et al. subsequently described the extended Kerr--BR geometry and studied its scalar-field mode
problem, including the associated two-horizon boundary conditions and outgoing spectral problem
\cite{Zhou:2026notes,Zhou:2026bridge}. These studies provide the
geometric and spectral framework for investigating wave propagation across the bridge.

Our previous work studied scalar scattering in the original exterior region, with a transparent boundary condition imposed at
$r\to+\infty$~\cite{Huang:2026transparent}. In that model, the neighbouring region is replaced by an auxiliary channel, and the resulting superradiant
range has a magnetic-field cutoff. The present work addresses a different question by retaining the neighbouring region and following the wave all the
way from the outer horizon to the inner horizon of that region. Instead of specifying an auxiliary response at the bridge, we impose wave conditions at
both horizon ends and calculate the complete bidirectional connection between them. This provides a direct way to determine how propagation through the
extended region changes the scattering response.

We study a neutral, minimally coupled massless scalar field and construct a $2\times2$ connection matrix for real frequencies. Its entries describe
reflection and transmission for waves incident from either horizon. The conserved radial flux then gives the condition for superradiant amplification in
terms of the two horizon frequencies and the transmission amplitude. We also examine the behaviour of the radial equation at the coordinate bridge and
continue its solutions through the neighbouring region. Since the extended geometry contains a ring singularity, we separately estimate the local scalar
energy on the positive-$r$ side of the ring.

The main results can be summarized as follows. First, when the neighbouring region is retained, transmission and superradiant amplification persist
beyond the magnetic-field cutoff found in the transparent-end model. The response is therefore determined by the full region included in the wave
problem, rather than by the local boundary condition at the bridge alone. Second, a negative value of the radial wave-number coefficient at the bridge
does not prevent transmission. A wave can cross a finite-width forbidden region and continue towards the second horizon. Third, for non-extremal
horizons, the amplification vanishes linearly as the frequency approaches either horizon threshold. Finally, the peaks in the real-frequency
amplification are associated with nearby damped resonances, identified through zeros of the analytically continued connection denominator. These results
are checked using two independent radial numerical implementations.

The connection matrix studied here describes a specific two-horizon, separated-mode problem. Its incoming waves are specified on two different
past-horizon branches, and its outgoing waves are defined relative to the corresponding future-horizon branches. The right endpoint is the inner horizon
of the neighbouring region, rather than a second asymptotically flat infinity. This distinction is important for interpreting the calculated fluxes.
Moreover, the extended geometry contains closed timelike curves, and the ring singularity requires a separate specification of the admissible
four-dimensional field configurations. We therefore focus on the stationary radial connection and its conserved flux, while examining the local field
behaviour near the ring independently. The relation between singularities, wave-operator domains and dynamical evolution has a broader mathematical
context
\cite{Wald:1980nonglobally,IshibashiWald:2003dynamics,
HorowitzMarolf:1995probes}.

The paper is organized as follows. Sections~\ref{sec:geometry} and~\ref{sec:separation} introduce the Kerr--BR geometry and derive the separated scalar
equations. Section~\ref{sec:bridge} describes the extension across the coordinate bridge and the local field behaviour near the ring.
Section~\ref{sec:channels} constructs the bidirectional connection matrix and derives the superradiant condition and the threshold expansions.
Section~\ref{sec:numerics} presents the numerical methods and their validation. The real-frequency response, its nearby complex-frequency resonances and
the role of the bridge as an interior scattering region are examined in Section~\ref{sec:results}. Section~\ref{sec:comparison} compares the retained
two-horizon problem with the transparent-end model. The final discussion and conclusions are given in Sections~\ref{sec:discussion} and
\ref{sec:conclusions}, respectively. Further details of the
separation, flux, horizon initialization and local energy estimates are provided in the appendices.

\section{Geometry and coordinate conventions}\label{sec:geometry}
We use signature $(-+++)$ and $G=c=\hbar=1$. The exact-metric coordinates and parameters are fixed by
\begin{align}
\dd s_g^2=\Omega^{-2}\bigg[&-\frac Q{\rho^2}(\dd t-a\sin^2\theta\,\dd\varphi)^2
+\frac{\rho^2}{Q}\dd r^2+\frac{\rho^2}{\widetilde P}\dd\theta^2\nonumber\\
&+\frac{\widetilde P\sin^2\theta}{\rho^2}
\big(a\dd t-(r^2+a^2)\dd\varphi\big)^2\bigg],\label{eq:metric}
\end{align}
where
\begin{gather}
\rho^2=r^2+a^2\cos^2\theta,\qquad H=r^2+a^2,\qquad
Q=(1+B^2r^2)\Delta,\qquad \Delta=\alpha r^2-2dr+a^2,\label{eq:functions}\\
\widetilde P=1+\beta\cos^2\theta,\qquad
\Omega^2=1+B^2r^2-B^2\Delta\cos^2\theta,\\
\begin{gathered}
I_1=1-\tfrac12B^2a^2,\quad I_2=1-B^2a^2,\quad d=MI_2/I_1,\\
\alpha=1-B^2M^2 I_2/I_1^2,\quad \beta=B^2(M^2I_2/I_1^2-a^2).
\end{gathered}\label{eq:parameters}
\end{gather}
These are the exact-solution conventions of ref.~\cite{Podolsky:2025tle}. The parameters $M$ and $a$ are not identified here with an ADM mass and a
specific ADM angular momentum. Kerr--BR has no asymptotically flat end at which that identification or a preferred normalization of $t$ is available.
Parameters such as $a/M$, $BM$ and $M\omega$ label calculations within the fixed metric convention and are not quantities measured against an established
asymptotic mass standard.

Axial regularity gives
\begin{equation}
\varphi\sim\varphi+2\pi C,\qquad C=(1+\beta)^{-1},\qquad
k=\frac nC,\quad n\in\mathbb Z.\label{eq:azimuth}
\end{equation}
The regular $2\pi$-periodic coordinate is $\phi=\varphi/C$. Its horizon angular velocity is $\varpi_j=\Omega_j/C$, so that $n\varpi_j=k\Omega_j$. We
restrict to $a\ne0$, $B\ne0$, $\alpha>0$, $\widetilde P>0$, and $d^2>\alpha a^2$, with the two positive roots
\begin{equation}
\rro=\frac{d+\sqrt{d^2-\alpha a^2}}{\alpha},\qquad
\rri=\frac{d-\sqrt{d^2-\alpha a^2}}{\alpha}.\label{eq:horizons}
\end{equation}
The relevant Killing-horizon quantities are
\begin{equation}
\Omega_j=\frac a{r_j^2+a^2},\qquad
\kappa_j=\frac{Q'(r_j)}{2(r_j^2+a^2)},\qquad j=\mathrm{o},\mathrm{i}.\label{eq:horizonquantities}
\end{equation}
With our choice of orientation, the inner-horizon surface gravity is negative, $\kappa_i<0$. The horizon generators are defined locally at each horizon.
The radial segment considered here extends from the black-hole outer horizon, across the coordinate bridge, to the inner horizon in the neighbouring
spacetime region.

\section{Separated scalar equations}\label{sec:separation}
Since the Maxwell stress tensor is traceless, the neutral minimally coupled massless equation is
\begin{equation}
\Box_g\Phi=(\Box_g-R[g]/6)\Phi=0.\label{eq:wave}
\end{equation}
For $\widetilde g=\Omega^2g$, the four-dimensional conformal identity is
\begin{equation}
(\Box_g-R[g]/6)(\Omega\Psi)=\Omega^3(\Box_{\widetilde g}-R[\widetilde g]/6)\Psi.
\end{equation}
It yields the separated field
\begin{equation}
\Phi=\Omega e^{-\ii\omega t+\ii k\varphi}R_{\ell n}(r)S_{\ell n}(\theta),\label{eq:field}
\end{equation}
with the required conformal factor~\cite{Gray:2025lwy}. Appendix~\ref{app:separation} gives the explicit operator and physical flux reduction.

Writing $x=\cos\theta$ and $p(x)=(1+\beta x^2)(1-x^2)$, the angular equation is
\begin{equation}
-\frac{\dd}{\dd x}\left(p\frac{\dd S}{\dd x}\right)
+\left[\frac{[k-a\omega(1-x^2)]^2}{p}+\frac13-\frac\beta3+2\beta x^2\right]S=\lambda S.\label{eq:angular}
\end{equation}
We impose $S\sim(1-x^2)^{|n|/2}$ at the poles and $\int_{-1}^{1}|S|^2\dd x=1$. The real-frequency eigenvalue is real. Its convention is fixed by the Kerr
limit
\begin{equation}
\lambda_{\ell n}=A_{\ell n}(a\omega)+\frac13-2an\omega+a^2\omega^2\quad(B=0),\label{eq:kerrlambda}
\end{equation}
where $A_{\ell n}$ is the scalar oblate-spheroidal eigenvalue~\cite{Berti:2005gp}.

The radial equation is
\begin{equation}
(QR')'+\left(\frac{K^2}{Q}+\frac{Q''}{6}-\lambda\right)R=0,
\qquad K=H\omega-ak.\label{eq:radial}
\end{equation}
With
\begin{equation}
u=\sqrt H\,R,\qquad \frac{\dd s}{\dd r}=\frac H Q,
\end{equation}
it becomes
\begin{align}
u_{ss}+V u&=0,\label{eq:schrodinger}\\
V&=\frac{K^2}{H^2}+\frac Q{H^2}\left(\frac{Q''}{6}-\lambda\right)
-\frac{Q(Q+rQ')}{H^3}+\frac{3r^2Q^2}{H^4}.\label{eq:potential}
\end{align}
The additive constant in $s$ is fixed at the bridge below. We call $V$ the radial wave-number coefficient: it depends on $\omega$ and on the angular
eigenvalue, and is not a frequency-independent mechanical potential.

\section{Extension across the coordinate bridge}
\label{sec:bridge}

To describe the geometry and wave propagation across the coordinate bridge at $r\to+\infty$, we introduce $y=1/r$, which places the bridge at $y=0$. We
first examine the regularity of the full physical metric, including the conformal factor, and then express the radial equation and physical scalar field
in this coordinate. This allows us to determine how the wave continues through the bridge and how the matching conditions follow from smoothness of the
physical field. We then consider the radial continuation towards the neighbouring inner horizon and the behaviour of the field near the ring singularity.

\subsection{Geometry and field continuation across the bridge}
\label{sec:bridge-geometry}

Set $y=1/r$ and define
\begin{equation}
\begin{gathered}
D=\alpha-2dy+a^2y^2,\qquad
F=(B^2+y^2)D,\qquad
A=1+a^2y^2,\\
\Sigma=1+a^2y^2\cos^2\theta,\qquad
W=B^2+y^2-B^2D\cos^2\theta.
\end{gathered}
\label{eq:bridge-definitions}
\end{equation}
The physical metric takes the exact form
\begin{align}
\dd s_g^2=\frac1W\bigg[&
-\frac F\Sigma(\dd t-a\sin^2\theta\dd\varphi)^2
+\frac\Sigma F\dd y^2
+\frac\Sigma{\widetilde P}\dd\theta^2\nonumber\\
&+\frac{\widetilde P\sin^2\theta}{\Sigma}
(ay^2\dd t-A\dd\varphi)^2
\bigg].
\label{eq:bridge-metric}
\end{align}

The non-extremal condition gives
\begin{equation}
d^2-\alpha a^2
=\frac{M^2I_2-a^2I_1^2}{I_1^2}>0,
\end{equation}
which implies $I_2>0$. Together with $B\ne0$ and $\alpha>0$, this gives $0<\alpha<1$. At the bridge, $y=0$, we have
\begin{equation}
F_0=B^2\alpha>0,\qquad
W_0=B^2(1-\alpha\cos^2\theta)>0.
\label{eq:bridge-limits}
\end{equation}
The metric determinant and the normal norm are
\begin{equation}
\det g=-\frac{\Sigma^2\sin^2\theta}{W^4},
\qquad
g^{yy}\big|_0
=B^4\alpha(1-\alpha\cos^2\theta)>0.
\label{eq:bridge-regular}
\end{equation}
After introducing regular angular charts on the axis, the bridge is analytic, nondegenerate and timelike. This local geometry agrees with the extension
in ref.~\cite{Zhou:2026bridge}. Since the bridge lies at finite tortoise distance, a wave can reach and cross it. The bridge therefore provides an
interior passage for the wave problem.

The radial coordinate and coefficient are also regular at $y=0$:
\begin{align}
\frac{\dd s}{\dd y}&=-\frac A F,\qquad s(0)=0,\\
V(y)&=\frac{(A\omega-ak y^2)^2}{A^2}
+\frac F{A^2}\left(\frac{F_{yy}}6-\lambda\right)
-\frac{a^2yFF_y}{A^3}
+\frac{a^2F^2(2A-3)}{A^4}.
\label{eq:regular-potential}
\end{align}
In particular, the value of the radial coefficient at the bridge is
\begin{equation}
q_{\mathcal B}^2\equiv V(0)
=\omega^2+B^2\alpha
\left[\frac{\alpha+B^2a^2}{3}-\lambda\right]
-a^2B^4\alpha^2.
\label{eq:bridge-value}
\end{equation}
Even when $q_{\mathcal B}^2=0$, the highest-derivative coefficient of the radial equation remains nonzero. Thus, the radial equation retains its regular
differential structure at this point.

The physical scalar field is represented locally by
\begin{equation}
\Phi=e^{-\ii\omega t+\ii k\varphi}
S(\theta)\sqrt{\frac WA}\,u(y).
\label{eq:regular-field}
\end{equation}
This expression gives the smooth continuation of the field through $y=0$. On both real coordinate charts, we use the positive roots
\begin{equation}
\Omega=\frac{\sqrt W}{|y|},
\qquad
\sqrt H=\frac{\sqrt A}{|y|}.
\end{equation}
Their common factor cancels in $\Omega/\sqrt H$, giving the smooth ratio $\sqrt{W/A}$ in Eq.\eqref{eq:regular-field}. We therefore continue the wave by
matching $u$ and its derivative $u_s$ across the bridge. This matching ensures continuity of the physical field and its radial derivative. The original
radial function $R(1/y)=|y|u/\sqrt A$ contains the nonsmooth factor $|y|$, so the continuation is specified through the regular field representative in
Eq.\eqref{eq:regular-field}.

\subsection{Radial continuation and the ring singularity}
\label{sec:ring-domain}

The selected radial interval starts at the outer horizon $r_o$, crosses the coordinate bridge from $r\to+\infty$ to $r\to-\infty$, and continues through
$r=0$ to the inner horizon $r_i$ of the neighbouring region. For $a\ne0$, $H(0)=Q(0)=a^2$, so the separated radial equation is regular at $r=0$. We
continue its solutions analytically through this point and use this continuation in the stationary calculations below.

The four-dimensional geometry, however, has a ring singularity at $r=0$, $\theta=\pi/2$, together with closed timelike curves in part of the neighbouring
region~\cite{Zhou:2026bridge,Carter:1968global}. The neighbouring region also contains isolated axial points where the conformal factor vanishes. On the
symmetry axis, these points are located at
\begin{equation}
r=r_\Omega=-\frac{I_1}{B^2M}
=-\frac{d}{1-\alpha}.
\label{eq:omega-zero-location}
\end{equation}
They form an additional axial boundary of the extended geometry and are associated with the null-infinity structure of the corresponding axial sections.
The four-dimensional field domain must therefore account for the behaviour of the scalar field at these locations.

The local behaviour of the scalar field near the ring requires a separate analysis. Each local analytic radial solution has the expansion
\begin{equation}
R=R_0+R_1r+O(r^2),
\label{eq:ring-radial-expansion}
\end{equation}
and the angular function is regular. Thus, the scalar field remains bounded near $r=x=0$, although its derivatives measured in a local orthonormal frame
can become unbounded.

A one-sided energy estimate can be obtained on the positive-$r$ side. Define
\begin{equation}
\mathcal L=\widetilde P H^2-a^2Q(1-x^2).
\label{eq:ring-L-definition}
\end{equation}
For sufficiently small $r>0$, the $t=\mathrm{constant}$ hypersurface is spacelike. Its unit normal $n^\mu$ and induced metric $h$ define the positive
local energy of the minimally coupled scalar:
\begin{equation}
E_+(\epsilon)=
\int_{\substack{0<\rho<\epsilon\\r>0}}
T_{\mu\nu}n^\mu n^\nu\,\dd\mathrm{vol}_h.
\label{eq:ring-energy}
\end{equation}
The integration includes the azimuthal period. Appendix~\ref{app:ring-energy} gives the corresponding local energy estimate for either independent
analytic radial solution:
\begin{equation}
T_{\mu\nu}n^\mu n^\nu\leq C_1\rho^{-2},
\qquad
E_+(\epsilon)\leq C_2\epsilon^{3/2}.
\label{eq:ring-integrability}
\end{equation}
The integrated energy therefore vanishes at least as fast as $\epsilon^{3/2}$ as the neighbourhood shrinks, even if the local energy density diverges.
The bound holds for arbitrary local radial data $(R_0,R_1)$, with constants that may depend on the solution and background. Finite energy on the
positive-$r$ side alone cannot select a unique boundary condition at the ring.

The energy estimate uses a constant-$t$ slice on the positive-$r$ side. Its causal character changes across the ring. For small negative $r$ at $x=0$,
the expansion of $\mathcal L$ gives
\begin{equation}
\mathcal L=2da^2r+O(r^2)<0.
\label{eq:ring-negative-r-L}
\end{equation}
The closed azimuthal curves then become timelike, and the constant-$t$ hypersurface is no longer spacelike. The positive spatial metric and timelike unit
normal used in Eq.\eqref{eq:ring-energy} therefore apply only to the positive-$r$ side.

The local energy on other spacelike hypersurfaces and the allowed field behaviour at the ring require further analysis. In particular, the relation
between the radial solutions on the two sides of the ring must be considered as part of the four-dimensional wave problem. These questions are important
for determining the admissible field configurations and their time
evolution~\cite{Wald:1980nonglobally,IshibashiWald:2003dynamics,HorowitzMarolf:1995probes}.

\section{Bidirectional real-frequency connection}\label{sec:channels}
\subsection{Horizon wave modes and the connection matrix}
We first define the incoming and outgoing wave modes at the outer horizon and the neighbouring inner horizon. For a fixed real frequency and angular
mode, a
\(2\times2\) connection matrix maps the incoming amplitudes at the two ends to the
outgoing amplitudes. Each column describes incidence from one end and gives the corresponding reflection and transmission amplitudes. On the selected
interval, $s\to-\infty$ at $r_o$ and $s\to+\infty$ at the neighbouring $r_i$. Define
\begin{equation}
\noo=\omega-k\Omega_o,\qquad \nii=\omega-k\Omega_i.\label{eq:nus}
\end{equation}
At each non-extremal endpoint $V\to\nu_j^2$. With the future-horizon orientations of ref.~\cite{Zhou:2026bridge}, the two outgoing Jost solutions are
\begin{equation}
f_L\sim e^{-\ii\noo s}\quad(s\to-\infty),\qquad
f_R\sim e^{+\ii\nii s}\quad(s\to+\infty).\label{eq:jost}
\end{equation}
For real \(\omega\) and fixed \(k\), we obtain the incoming radial solutions by complex conjugating the outgoing ones. The time and azimuthal factors
remain unchanged, so the horizon frequencies \(\nu_j\) also remain unchanged. In the chosen segment, the right endpoint lies at the neighbouring inner
horizon, on the side leading towards a white-hole-like region; it is not a second asymptotically flat infinity. Near either horizon, a wave with radial
momentum \(p\) carries current
\(p|u|^2\), while its propagation direction is determined by the sign of \(p/\nu_j\).
We therefore retain the sign of \(\nu_j\) in Eq.\eqref{eq:jost}: replacing it with
\(|\nu_j|\) would interchange the incoming and outgoing assignments when \(\nu_j<0\).

For incidence from the left and right, respectively, define all four amplitudes by
\begin{align}
u_L&\sim\begin{cases}e^{+\ii\noo s}+R_L e^{-\ii\noo s},&s\to-\infty,\\T_L e^{+\ii\nii s},&s\to+\infty,\end{cases}\label{eq:left-incident}\\
u_R&\sim\begin{cases}T_R e^{-\ii\noo s},&s\to-\infty,\\e^{-\ii\nii s}+R_R e^{+\ii\nii s},&s\to+\infty.\end{cases}\label{eq:right-incident}
\end{align}
The endpoint amplitude is normalized to one. This avoids imposing the divergent factor $|\nu_j|^{-1/2}$ close to a horizon threshold.

The conserved radial current is
\begin{equation}
J=\operatorname{Im}(u^*u_s)=Q\operatorname{Im}(R^*R'),
\qquad \mathcal F=2\pi C J.\label{eq:current}
\end{equation}
This Wronskian form of the radial flux is the standard basis for relating endpoint amplitudes in separated black-hole wave
equations~\cite{Chandrasekhar:1983mathematical}. It gives
\begin{equation}
|R_L|^2+\frac{\nii}{\noo}|T_L|^2=1,\qquad
|R_R|^2+\frac{\noo}{\nii}|T_R|^2=1,\label{eq:flux}
\end{equation}
\begin{equation}
\nii T_L=\noo T_R, \qquad |R_L|=|R_R|.\label{eq:reciprocity}
\end{equation}
Writing the incoming amplitudes as $a=(a_L,a_R)^T$ and the outgoing amplitudes as $b=(b_L,b_R)^T$, the two-horizon connection matrix is defined by
$b=\mathsf S a$. Equivalently,
\begin{equation}
\mathsf S=\begin{pmatrix}R_L&T_R\\T_L&R_R\end{pmatrix},\qquad
\mathsf S^\dagger\eta\mathsf S=\eta,\qquad \eta=\operatorname{diag}(\noo,\nii).\label{eq:pseudounitary}
\end{equation}
When \(\nu_o\) and \(\nu_i\) have opposite signs, the flux balance contains both positive and negative contributions, so it cannot be interpreted as a
sum of nonnegative probabilities. With the chosen future-directed horizon generator, a channel is called positive norm for \(\nu_j>0\) and negative norm
for \(\nu_j<0\). These labels describe its local Klein–Gordon signature, rather than a Hilbert-space norm on a global Cauchy surface. For real
\(\omega\), nonzero \(\nu_o\) and
\(\nu_i\), and a finite connection matrix, Eq.\eqref{eq:flux} gives
\begin{equation}
Z_L\equiv |R_L|^2-1=-\frac{\nii}{\noo}|T_L|^2.
\end{equation}
For \(T_L\ne0\), \(Z_L>0\) holds if and only if \(\nu_o\nu_i<0\). This requires opposite signs but allows either ordering. A positive-norm wave incident
from the left has \(\nu_o>0\), so amplification then requires \(\nu_i<0\). The amplification criterion is therefore
\begin{equation}
\noo>0,\qquad \nii<0,\qquad T_L\ne0.
\end{equation}
For $k>0$ and $\Omega_i>\Omega_o$, the two sign conditions are equivalent to
\begin{equation}
k\Omega_o<\omega<k\Omega_i,\label{eq:gainband}
\end{equation}
In this band, amplification requires a finite connection matrix and \(T_L\ne0\). The two columns give the responses to a positive-norm wave incident from
the left and a negative-norm wave incident from the right. We use \(Z\equiv Z_L\) for the gain under left incidence, calculated from the conserved flux
with the channel signs retained. This gain characterizes the chosen separated-mode continuation; by itself, it does not define an observer-independent
scattering cross section.

\subsection{Past and future horizon data}\label{sec:horizon-data}
The two columns of \(\mathsf S\) describe the responses to incoming waves on two different past-horizon branches: one at the original outer horizon and
the other at the neighbouring inner horizon. The two inputs are specified on their respective horizon branches, rather than at a common asymptotically
flat infinity. We identify the two branches using only the local geometry near the horizons. Define
\begin{equation}
w_-=t-s,\qquad w_+=t+s,\qquad
\psi(r)=\int^r\frac{a}{Q(\widetilde r)}\dd\widetilde r.
\end{equation}
Near either simple horizon, $\psi-\Omega_js$ is smooth. The regular advanced and retarded angles are $\varphi_v=\varphi+\psi$ and
$\varphi_u=\varphi-\psi$. Their co-rotating versions are $\phi_j^v=\varphi_v-\Omega_jw_+$ and $\phi_j^u=\varphi_u-\Omega_jw_-$. Up to a smooth radial
phase, a $v$-type mode has phase $e^{-\ii\nu_jw_++\ii k\phi_j^v}$ and a $u$-type mode has phase $e^{-\ii\nu_jw_-+\ii k\phi_j^u}$.
Appendix~\ref{app:frobenius} gives the phase identities explicitly.

Use the same stationary-time convention at both ends and choose $\chi_j=\partial_t+\Omega_j\partial_\varphi$ to be future-directed in each local
stationary horizon collar. At fixed co-rotating angle, increasing $t$ follows $\chi_j$. The corresponding local Kruskal coordinates are
\begin{align}
U_o=-e^{-\kappa_ow_-},\quad &V_o=e^{\kappa_ow_+},\qquad U_o<0<V_o,\\
U_i=+e^{-\kappa_iw_-},\quad &V_i=-e^{\kappa_iw_+},\qquad V_i<0<U_i.\label{eq:local-kruskal}
\end{align}
We retain the signed surface gravities \(\kappa_o>0\) and \(\kappa_i<0\). Near either horizon, \(U_jV_j=-e^{2\kappa_js}\) vanishes linearly in the radial
coordinate difference \(|r-r_j|\). Inside the coordinate regions defined above, both
\(U_j\) and \(V_j\) increase along \(\chi_j\), since \(\chi_jU_j=-\kappa_jU_j>0\)
and \(\chi_jV_j=\kappa_jV_j>0\). These coordinates therefore follow the same local future convention at both ends, without requiring \(\partial_t\) to be
timelike throughout the retained region. The corresponding past and future horizon branches are listed in table~\ref{tab:horizon-data}.

\begin{table}[htbp]
\centering\small
\begin{tabular}{cccc}
\toprule
Endpoint & Radial wave & Phase type & Horizon data relative to the segment\\
\midrule
Left & $\overline f_L\sim e^{+\ii\nu_os}$ & $u$ & $\mathcal H_o^-:V_o=0$, incoming\\
Left & $f_L\sim e^{-\ii\nu_os}$ & $v$ & $\mathcal H_o^+:U_o=0$, outgoing\\
Right & $\overline f_R\sim e^{-\ii\nu_is}$ & $v$ & $\mathcal H_i^-:U_i=0$, incoming\\
Right & $f_R\sim e^{+\ii\nu_is}$ & $u$ & $\mathcal H_i^+:V_i=0$, outgoing\\
\bottomrule
\end{tabular}
\caption{Local past and future horizon branches for the specified two-horizon segment. Incoming and outgoing mean entering and leaving that segment. The barred symbols conjugate the radial solution at fixed $(\omega,k)$.}
\label{tab:horizon-data}
\end{table}

We consider monochromatic modes with a fixed real frequency \(\omega\). The left outgoing wave enters the original black-hole horizon. The right outgoing
wave leaves the selected segment through the neighbouring inner horizon towards a white-hole-like region~\cite{Zhou:2026bridge}. This right horizon
branch is the future boundary of the local stationary region used here. It represents passage out of the segment, rather than ordinary black-hole
absorption or detection at an asymptotically flat infinity. The past branches specify the two inputs to
\(\mathsf S\).

The region contains closed timelike curves, so the two horizon inputs cannot be treated as ordinary initial data on a single surface for the whole
region. The field near the ring singularity also needs further study before a well-defined time evolution from these inputs can be established. In the
amplification band, the left input has positive norm and the right input has negative norm. Keeping these signs, we calculate \(Z\) as the flux gain for
the chosen radial continuation. To connect this result with an experiment, we would need a region with well-defined wave evolution, as well as a clear
description of how the incoming waves are generated and the outgoing waves are measured.

\subsection{Coordinate-independent channel labels}
We use labels that remain unchanged under a rescaling of stationary time or a shift to a uniformly rotating frame. These labels describe the signs of the
horizon channels and the frequency's position relative to the two horizon thresholds.

Consider the transformation $t'=c_t t$, $\varphi'=\varphi-\varpi_0t$, where $c_t>0$. The frequencies transform as
\begin{equation}
\omega'=\frac{\omega-k\varpi_0}{c_t},
\qquad
\Omega_j'=\frac{\Omega_j-\varpi_0}{c_t},
\qquad
\nu_j'=\frac{\nu_j}{c_t}.
\label{eq:frequency-transform}
\end{equation}
Since $c_t>0$, the sign of each $\nu_j$, and hence the sign of $\nu_o\nu_i$, is preserved. The ratio $\nu_o/\nu_i$ is also unchanged when $\nu_i\ne0$. A
useful invariant frequency label, defined for $k(\Omega_i-\Omega_o)\ne0$, is
\begin{equation}
\xi=\frac{\omega-k\Omega_o}{k(\Omega_i-\Omega_o)}.
\label{eq:xi}
\end{equation}
The values $\xi=0$ and $\xi=1$ mark the outer- and inner-horizon thresholds, respectively. For $k>0$ and $\Omega_i>\Omega_o$, the conditions $\nu_o>0$
and $\nu_i<0$ become $0<\xi<1$.

With consistent normalization of the endpoint waves, the gain $Z$ is unchanged as well. But the phases of individual reflection and transmission
amplitudes still depend on the phase convention used for the horizon Jost solutions.

\subsection{Single-horizon thresholds}

We study the limit in which one horizon frequency approaches zero while the other remains nonzero. For non-extremal horizons and a smooth, nondegenerate
angular eigenvalue branch, the gain $Z$ vanishes linearly with the horizon frequency approaching zero. We derive this behaviour by expanding and matching
the radial solutions near the corresponding endpoint.

First consider the outer-horizon threshold $\noo=0$, with $\nii\ne0$. The right-outgoing solution has the limiting form $f_{R,0}=a_o+b_o s+o(1)$ as
$s\to-\infty$. Current conservation gives $\operatorname{Im}(a_o^*b_o)=\nii$, so $b_o\ne0$. At this threshold, the two left-end wave factors
$e^{\pm\ii\noo s}$ coincide. Matching their expansions for small nonzero $\noo$ to the limiting solution gives
\begin{equation}
T_L=\frac{2\ii\noo}{b_o}+O(\noo^2),
\qquad
Z=-\frac{4\nii\noo}{|b_o|^2}+O(\noo^2).
\label{eq:outer-threshold}
\end{equation}

At the inner-horizon threshold $\nii=0$, with $\noo\ne0$, the left-outgoing solution has the limiting form $f_{L,0}=a_i+b_i s+o(1)$ as $s\to+\infty$. Its
current gives $\operatorname{Im}(a_i^*b_i)=-\noo$, so $b_i\ne0$. Applying the same matching procedure at the right endpoint and using $|R_L|=|R_R|$
yields
\begin{equation}
Z=-\frac{4\noo\nii}{|b_i|^2}+O(\nii^2).
\label{eq:inner-threshold}
\end{equation}

These limits are set by the vanishing of a horizon frequency. The condition $\qb^2=0$ at the bridge marks a zero of the local radial coefficient at a
regular interior point. That condition alone does not define a horizon channel threshold.

\section{Numerical construction}\label{sec:numerics}

For each background and real frequency, we first solve the angular eigenvalue problem, then propagate the two horizon solutions to a common point and
extract the reflection and transmission amplitudes. We check the angular approximation, radial propagation and amplitude extraction separately. We use
two radial numerical implementations to compare the full calculation.

\paragraph{Angular eigenvalue.}
We expand the angular eigenfunction in associated Legendre functions with the required parity and regular behaviour at both poles. Projecting the angular
equation onto this basis gives a symmetric generalized eigenvalue problem, whose matrix elements are evaluated by Gauss quadrature.

We check angular convergence by increasing the number of basis functions. We also reconstruct the angular eigenfunction, substitute it into the original
differential equation, and evaluate the residual on a separate, denser quadrature grid. This checks how well the numerical function satisfies the
differential equation, in addition to the matrix eigenvalue problem. At \(B=0\), we check our angular eigenvalues against independent oblate-spheroidal
results, using Eq.\eqref{eq:kerrlambda} to convert the reference values to our definition of \(\lambda\)~\cite{Berti:2005gp}.

\paragraph{Horizon initialization.}
We initialize the outgoing solutions a short distance from the horizons, at $r_o+\epsilon$ and $r_i-\epsilon$, using second-order Frobenius expansions.
These expansions impose the required outgoing behaviour and provide starting values for both $u$ and $u_s$. Their exponents and coefficients are given in
appendix~\ref{app:frobenius}.

\paragraph{Radial propagation.}
We integrate the radial equation as a first-order system for the state $(u,u_s)$, using the same tortoise derivative in every coordinate chart. The main
implementation uses $r$ away from the bridge, switches to $y=1/r$ near the bridge, and returns to the neighbouring $r$ chart after crossing it. Passing
the same state between charts preserves the continuation specified in section~\ref{sec:bridge}.

The second implementation uses a continuous compact coordinate,
\begin{equation}
\begin{aligned}
&\frac{r}{M}=\tan\chi,
\qquad \chi_{\mathcal B}=\frac{\pi}{2},\\
&\arctan(r_o/M)<\chi<\pi+\arctan(r_i/M).
\end{aligned}
\label{eq:compact-radial-chart}
\end{equation}
Here the bridge lies at a finite coordinate value and can be crossed directly. This implementation evaluates $V$ from the homogeneous polynomial
representation of $H^4V$, with the cancellations among the high-order terms performed analytically before numerical evaluation. The resulting expression
can be evaluated at the bridge without subtracting individually divergent terms.

\paragraph{Amplitude extraction.}
The two outgoing solutions are transported to the same geometric matching point. In the main implementation, this is the neighbouring $r=0$ section,
using the analytic radial continuation specified above. At that point, define the solution states $\mathbf f_L=(f_L,\partial_s f_L)^T$ and
$\mathbf f_R=(f_R,\partial_s f_R)^T$. For real frequency, their radial complex conjugates provide the incoming states. The reflection and transmission
amplitudes follow from
\begin{equation}
[\mathbf f_L,-\mathbf f_R]
\binom{R_L}{T_L}
=-\overline{\mathbf f_L},
\qquad
[\mathbf f_L,-\mathbf f_R]
\binom{T_R}{R_R}
=\overline{\mathbf f_R}.
\label{eq:match}
\end{equation}
Solving these two systems gives the two columns of $\mathsf S$. The Wronskian formulas in appendix~\ref{app:frobenius} provide a second extraction of the
amplitudes from the same propagated solutions.

\paragraph{Numerical validation.}
We compare the two radial implementations at identical parameters and check amplitude extraction with the Wronskian formulas. Further checks use
signed-flux conservation, reciprocity, Eq.\eqref{eq:pseudounitary} and the denominator diagnostics in appendix~\ref{app:denominator}. We test convergence
by varying the angular resolution, horizon starting distance, integration tolerances and matching point while keeping the horizon conditions fixed. The
threshold behaviour is checked against Eqs.\eqref{eq:outer-threshold} and \eqref{eq:inner-threshold}.

\section{Real-frequency response}\label{sec:results}
\subsection{Frequency response of the dipole mode}

We set $M=1$, $a/M=0.9$ and $\ell=n=1$, and compare $BM=0.23$, $BM_{\mathrm{ref}}$, $0.26$ and $0.30$. \(\xi\) and \(Z_{\rm peak}\) represent the
dimensionless frequency position and the maximum superradiant amplification, respectively. The reference case $BM_{\mathrm{ref}}$ corresponds to the
boundary of the transparent-end model, imposed at the coordinate endpoint $r\to+\infty$. These backgrounds probe the retained two-horizon response below,
near and above that cutoff.

For each background, we scan $0.01\le\xi\le0.99$. Table~\ref{tab:peaks} lists the horizon thresholds and the resulting peak values.

\begin{table}[htbp]
\centering\small
\begin{tabular}{cccccc}
\toprule
$BM$ & $Mk\Omega_o$ & $Mk\Omega_i$ & $\xi_{\rm peak}$ &
$M\omega_{\rm peak}$ & $100Z_{\rm peak}$\\
\midrule
0.23 & 0.299304 & 0.798505 & 0.159347 & 0.378850 & 21.17410\\
$BM_{\mathrm{ref}}$ & 0.297678 & 0.798523 & 0.190104 & 0.392890 & 11.33627\\
0.26 & 0.295341 & 0.798519 & 0.228629 & 0.410382 & 5.61456\\
0.30 & 0.289269 & 0.798344 & 0.308538 & 0.446338 & 1.73772\\
\bottomrule
\end{tabular}
\caption{Horizon thresholds and resolved dipole peaks for
$a/M=0.9$ and $\ell=n=1$.
$BM_{\mathrm{ref}}=0.2427791435$.
The last column gives the amplification in percent.}
\label{tab:peaks}
\end{table}

\begin{figure}[htbp]
\centering
\includegraphics[width=0.616\textwidth]{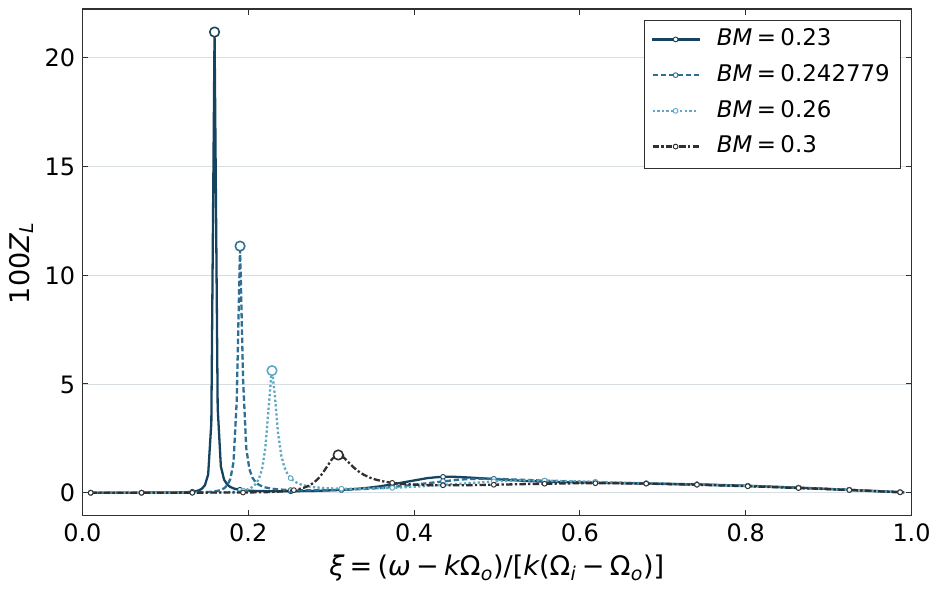}
\caption{Dipole amplification for the four backgrounds in
table~\ref{tab:peaks}. The curves cover $0.01\le\xi\le0.99$.
The horizon thresholds lie at $\xi=0$ and $\xi=1$.}
\label{fig:gain}
\end{figure}

Figure~\ref{fig:gain} shows finite positive amplification at the reference field and at the larger fields $BM=0.26$ and $0.30$. The retained two-horizon
channel therefore continues to amplify beyond the transparent-end cutoff $r\to+\infty$. Across these four backgrounds, the peak height decreases as $BM$
increases, while its position shifts to larger $\xi$. The two horizon thresholds remain separated throughout the sample.

The sign of the gain is determined by the relative signs of the two horizon frequencies, while its magnitude is governed by their frequency ratio and the
transmission amplitude. The latter is determined by propagation through the full radial coefficient, including its dependence on the angular eigenvalue.

\subsection{Real-frequency denominator and conditioning}
\label{sec:denominator}
We first study the response at real frequencies because our main interest is the reflection, transmission and superradiant amplification of incident
waves. These quantities describe the scattering response to waves with a specified real frequency. Complex-frequency analysis addresses a different
question: it identifies outgoing modes and their resonance frequencies. We therefore use the real-frequency response to study the amplification and its
dependence on the background parameters, and examine the nearby complex-frequency poles separately to understand the origin of the response peaks.

Flux conservation alone cannot exclude a real zero of $\mathcal D=W(f_L,f_R)$ in the mixed-sign interval. We therefore examine the denominator and the
conditioning of the matching matrix throughout the frequency band. The numerical definitions and methods are given in Appendix~\ref{app:denominator}.

\begin{figure}[htbp]
\centering
\includegraphics[width=0.616\textwidth]{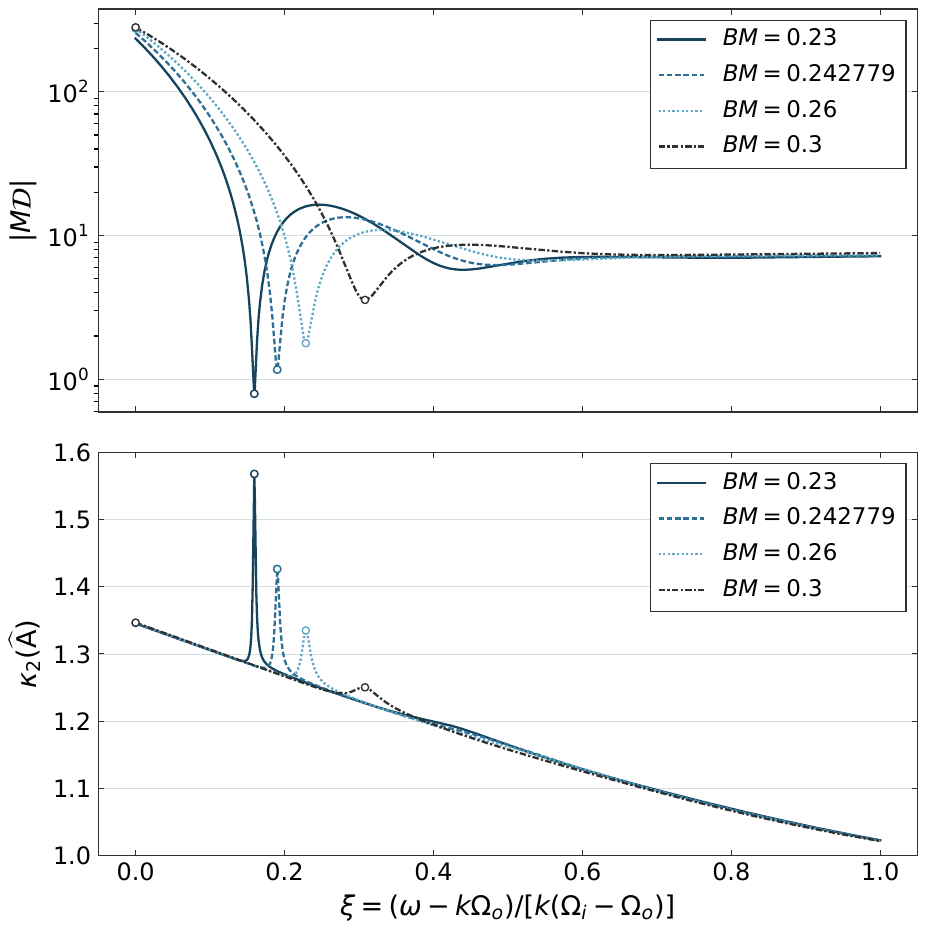}
\caption{Real-frequency denominator magnitude and condition number of the column-normalized matching matrix for the four dipole backgrounds. The curves show how the denominator and matrix conditioning vary across the frequency band, including the locations of the denominator minima.}
\label{fig:denominator}
\end{figure}

Table~\ref{tab:denominator} and Fig.~\ref{fig:denominator} summarize the results. No real zero of $\mathcal D$ is found in our numerical calculations. At
the sampled and refined frequencies on the four backgrounds, the column-normalized matching matrix satisfies $\kappa_2(\widehat{\mathsf A})<1.568$.
Independent numerical calculations give consistent results, supporting the reliability of the reported denominator minima. These results also show that
the amplification peaks are not caused by a nearly singular matching matrix at real frequencies.

At the exact horizon thresholds, the analytic expansions give $\mathcal D\to b_o\ne0$ or $\mathcal D\to-b_i\ne0$. Thus, for each fixed background, the
denominator remains nonzero sufficiently close to either endpoint. However, the numerical results cannot exclude a narrow minimum or a real zero between
the thresholds. Therefore, no real zero is found in our calculations, but we cannot prove that the denominator remains nonzero throughout the entire
continuous frequency interval.

\begin{table}[htbp]
\centering\small
\begin{tabular}{cccccc}
\toprule
$BM$ & $\xi_{\min|M\mathcal D|}$ & $\min|M\mathcal D|$ &
$\min\widehat D$ &
$\max\kappa_2(\widehat{\mathsf A})$ &
$\max\kappa_2(\mathsf A)$\\
\midrule
0.23 & 0.1593392 & 0.794106 & 0.906720 & 1.56799 & 139.19\\
$BM_{\mathrm{ref}}$ & 0.1900828 & 1.167345 & 0.940080 & 1.42643 & 152.00\\
0.26 & 0.2285614 & 1.783481 & 0.957457 & 1.34583 & 161.66\\
0.30 & 0.3080731 & 3.566733 & 0.957309 & 1.34655 & 162.87\\
\bottomrule
\end{tabular}
\caption{Extrema found in the real-frequency scans using the state $(u,Mu_s)$ at $r=0$. The second column gives the location of the minimum of the raw denominator $|M\mathcal D|$, which differs from that of the normalized denominator $\widehat D$. For the first two backgrounds, the normalized minima occur at $\xi=0.1594017$ and $0.1902046$, respectively, while for the last two backgrounds they occur at the smallest sampled values of $\xi$. The large raw condition numbers near the left endpoint are mainly due to the unequal magnitudes of the two matrix columns. After normalizing each column to unit length, the condition number no longer reflects the difference in column magnitudes.}
\label{tab:denominator}
\end{table}

\subsection{Local spectral analysis of the response peaks}
\label{sec:poles}

The peaks in the real-frequency response suggest a connection with nearby complex-frequency resonances. To examine this connection, we continue the
denominator $\mathcal D=W(f_L,f_R)$ into the complex-frequency plane and solve the outgoing condition $\mathcal D(\omega_p)=0$. For each of the four
dipole backgrounds, we find a nearby simple zero in the lower half-plane. Its position is listed in table~\ref{tab:poles}. With the time dependence
$e^{-\ii\omega t}$, a negative imaginary part corresponds to damping. These roots provide a spectral interpretation of the real-frequency peaks within
the selected two-horizon problem.

\begin{table}[htbp]
\centering\small
\begin{tabular}{cccc}
\toprule
$BM$ & $\operatorname{Re}(M\omega_p)$ &
$-\operatorname{Im}(M\omega_p)$ &
$|\dd(M\mathcal D)/\dd(M\omega)|$\\
\midrule
0.23 & 0.378832990 & 0.000855900 & 930.1\\
$BM_{\mathrm{ref}}$ & 0.392832905 & 0.001616718 & 724.4\\
0.26 & 0.410163320 & 0.003307449 & 540.9\\
0.30 & 0.444337689 & 0.011089389 & 321.2\\
\bottomrule
\end{tabular}
\caption{Numerically located simple poles near the dipole response
peaks. The derivative is evaluated using the endpoint
normalization of Appendix~\ref{app:frobenius}. Its nonzero value
confirms the local simple-root behaviour. }
\label{tab:poles}
\end{table}

The roots remain stable under changes in numerical resolution, horizon initialization and radial integration accuracy (Appendix~\ref{app:denominator}).
An independent calculation using a compact radial coordinate gives $|M\mathcal D|\lesssim1.1\times10^{-9}$ at the unrounded candidate roots. One further
Newton correction reduces the residuals below $10^{-13}$. Together with the nonzero derivatives and the convergence tests, these results support the
identification of four simple roots. An independent finite-difference calculation at $BM=0.23$ gives
$\dd(M\mathcal D)/\dd(M\omega)\simeq885.81-283.74\ii$, in agreement with the main calculation. The reported numerical precision is consistent with the
accuracy established by these independent checks.

For complex frequencies, the incoming left Jost solution is defined by analytically continuing $f_L^{\rm in}\sim e^{+\ii\nu_o s}$ from the real-frequency
axis. This continuation preserves the incoming-wave boundary condition. The solution is not obtained by complex conjugating $f_L$ at fixed complex
$\omega$, since complex conjugation would change the radial equation. With this analytically continued branch, the transmission identity
$T_L=2\ii\nu_o/\mathcal D$ remains valid. Near a simple zero with nonzero horizon frequency shifts, the transmission amplitude takes the form
\begin{equation}
T_L(\omega)=
\frac{2\ii\nu_o(\omega_p)}
{\mathcal D'(\omega_p)(\omega-\omega_p)}
+O(1).
\label{eq:pole-local}
\end{equation}
Here $\mathcal D'=\dd\mathcal D/\dd\omega$ is the total derivative along the selected angular eigenvalue branch $\lambda(\omega)$, rather than a
derivative at fixed $\lambda$.

The pole contributes a resonant term to the connection matrix. Since $\operatorname{Im}\omega_p<0$, the resonance is damped and the real-frequency
response remains finite. The maximum of the real-frequency gain need not occur exactly at $\operatorname{Re}\omega_p$, because both the numerator and the
regular part of the response vary with frequency.

Across the four backgrounds, the poles move farther into the lower half-plane as the real-frequency gain peaks become weaker. This behaviour shows that
the damping of the nearby resonances is closely related to the strength of the real-frequency response. The four poles thus provide a consistent spectral
interpretation of the amplification peaks in the two-horizon connection.

\subsection{The bridge as an interior scattering region}

The radial equation
\begin{equation}
u_{ss}+V(s)u=0
\end{equation}
shows how the radial propagation of the wave depends on the coefficient $V(s)$. When $V>0$, the local solutions have oscillatory behaviour, corresponding
to a propagating region. When $V<0$, they have exponential behaviour, indicating a locally forbidden region. The point $V=0$ marks the transition between
these two local behaviours. However, a region with $V<0$ does not necessarily prevent transmission through the full radial interval. A wave can still
cross a finite potential barrier and reach the other side.

\begin{figure}[htbp]
\centering
\includegraphics[width=0.616\textwidth]{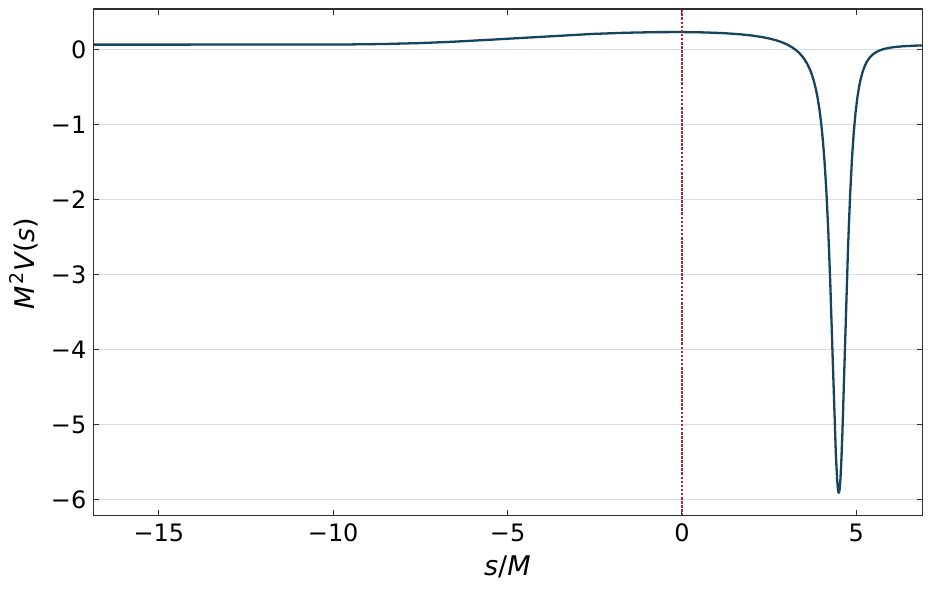}
\caption{Radial coefficient $M^2V$ across the coordinate bridge
as a function of the tortoise coordinate $s/M$, with the bridge
at $s=0$. The parameters are $BM_{\mathrm{ref}}$,
$a/M=0.9$, $\ell=n=1$ and $\xi=1/2$, corresponding to $M\omega=0.548100186$. The two horizons lie at
$s\to\mp\infty$. The coefficient is regular at the bridge
and along the separated radial continuation through $r=0$.}
\label{fig:potential}
\end{figure}

To show the radial structure across the entire bridge, we plot $V$ against the continuous tortoise coordinate $s$, with the bridge located at $s=0$. We
choose $\xi=1/2$, corresponding to the midpoint of the superradiant frequency interval. For the chosen background, this gives $M\omega=0.548100186$. As
shown in Fig.~\ref{fig:potential}, the tortoise coordinate covers both sides of the bridge and the neighbouring $r=0$ section in a single description.
The radial coefficient approaches the horizon values $\nu_o^2$ and $\nu_i^2$ at the two ends, respectively. The figure illustrates how the local
propagation conditions change from one horizon to the other, with sign changes of $V$ indicating local potential barriers in the interior, while both
horizon channels remain well defined.

We now examine whether a locally forbidden region at the bridge prevents transmission. For $BM=0.30$, the bridge value vanishes at
\begin{equation}
M\omega_0=0.3566065944,
\qquad V(0)=0.
\end{equation}
This frequency lies inside the two-horizon amplification interval. Table~\ref{tab:bridgezero} shows the response at frequencies on both sides of this
point. As $V(0)$ changes from negative to positive, the gain remains smooth and nonzero. In particular, when $V(0)<0$, the transmitted amplitude is still
finite, with $|T_L|^2=2.19126\times10^{-6}$ at the lowest frequency listed in the table. The two independent numerical implementations also give
consistent values of $Z$ at all three frequencies.

These results show that a negative value of the radial coefficient at the bridge does not close the transmission channel. The bridge acts as an interior
scattering region: local potential barriers affect the transmission, but do not by themselves prevent a wave from crossing the full two-horizon interval.
Thus, $V(0)>0$ is not a necessary condition for transmission.

\begin{table}[htbp]
\centering
\begin{tabular}{ccc}
\toprule
$M\omega$ & $M^2V(0)$ & $Z$\\
\midrule
0.3556065944 & $-8.239351\times10^{-4}$ & $1.4624433\times10^{-5}$\\
0.3566065944 & $0$ & $1.5478892\times10^{-5}$\\
0.3576065944 & $+8.258294\times10^{-4}$ & $1.6385347\times10^{-5}$\\
\bottomrule
\end{tabular}
\caption{Scattering response across the bridge coefficient zero
at $BM=0.30$, $a/M=0.9$ and $\ell=n=1$.}
\label{tab:bridgezero}
\end{table}

\subsection{Superradiant Gain near the Horizon Thresholds}

The amplification vanishes as the frequency approaches either horizon threshold. Eqs.\eqref{eq:outer-threshold} and
\eqref{eq:inner-threshold} predict a linear dependence of the
gain on the corresponding horizon frequency shift. To examine this behaviour, we consider the dipole mode at $BM_{\mathrm{ref}}$. The two sequences
approach their respective thresholds from inside the amplification interval, where $\nu_o>0$ at the left threshold and $\nu_i<0$ at the right threshold.
As shown in table~\ref{tab:thresholds}, the ratios $Z/(M\nu_j)$ approach finite, nonzero limits. Their opposite signs follow from the opposite signs of
the two horizon frequency shifts.

\begin{table}[htbp]
\centering
\begin{tabular}{ccc}
\toprule
$|M\nu_j|$ & $Z/(M\nu_o)$, left threshold &
$Z/(M\nu_i)$, right threshold\\
\midrule
$10^{-4}$ & $3.048313\times10^{-5}$ & $-3.832386\times10^{-2}$\\
$5\times10^{-5}$ & $3.042476\times10^{-5}$ & $-3.832696\times10^{-2}$\\
$2.5\times10^{-5}$ & $3.039563\times10^{-5}$ & $-3.832851\times10^{-2}$\\
Zero-shift Jost result & $3.036672\times10^{-5}$ & $-3.833014\times10^{-2}$\\
\bottomrule
\end{tabular}
\caption{Gain slopes near the two horizon thresholds for the
dipole mode at $a/M=0.9$ and $BM=0.2427791435$. The last row
gives independent estimates from the zero-shift Jost solutions
and their derivative coefficients $b_j$, with horizon cutoff $\epsilon=10^{-7}M$.}
\label{tab:thresholds}
\end{table}

The zero-shift Jost solutions at the opposite horizons provide an independent way to determine the limiting slopes. By extracting the derivative
coefficient $b_j$, we obtain slope estimates that agree with the limits approached by the finite-shift scattering results. The estimates also remain
stable as the endpoint cutoff is reduced, supporting the numerical evaluation of the linear coefficients in the analytic threshold expansions. For
example, reducing the cutoff from $10^{-6}M$ to $10^{-7}M$ changes the left slope from $3.036811\times10^{-5}$ to $3.036672\times10^{-5}$, and the right
slope from $-3.833072\times10^{-2}$ to $-3.833014\times10^{-2}$. This calculation focuses on the two horizon channels and uses the zero-shift Jost
solutions to determine the threshold slopes.

\section{Transparent-End Model}
\label{sec:comparison}

The transparent-end model at $r\to+\infty$ considers the original exterior $r_o<r<+\infty$ and attaches an auxiliary constant-coefficient channel at the
coordinate endpoint, which corresponds to $s=0$. When $q_c^2=V(0)>0$, the two basis waves in this channel satisfy
\begin{equation}
u_s=\pm\ii q_cu
\end{equation}
at the matching point. These relations define a Robin boundary condition for the exterior region. An incident wave from the auxiliary channel is
amplified when $0<\omega<k\Omega_o$ and $q_c^2>0$. For $a/M=0.9$ and $\ell=n=1$, the two conditions cease to overlap at $BM\simeq0.243$. This value marks
the end of the amplification range in the transparent-end model.

The scattering response is different when the neighbouring region beyond $r\to+\infty$ is retained and the wave propagates through the full interval
between the two horizons. At $BM=0.30$, the results in section~\ref{sec:results} show finite transmission and gain on both sides of the bridge condition
$q_{\mathcal B}^2=0$. This behaviour follows from the finite width of the locally forbidden region: a wave can cross the barrier by tunnelling and
continue towards the other horizon. Thus, the local sign of the radial coefficient at the bridge does not determine whether transmission through the full
two-horizon region is possible.

The role of the matching point can be understood through the response function $Y=u_s/u$. For a fixed exterior solution, it satisfies
\begin{equation}
Y_s=-V-Y^2.
\label{eq:impedance}
\end{equation}
Propagating $Y$ according to this equation, or propagating the full state vector when $u=0$, preserves the same exterior solution and scattering problem.
Assigning a new value $Y=+\ii\sqrt{V(s_c)}$ at each matching point $s_c$, however, imposes a new boundary response. This changes the auxiliary exterior
and defines a different scattering model. The transfer law in Appendix~\ref{app:cut} describes how the response evolves between matching points when the
exterior solution is kept fixed.

The two models therefore describe different scattering problems, with different exterior regions and boundary conditions. The value $BM_{\mathrm{ref}}$
characterizes the amplification range of the specified transparent-end model, while the retained two-horizon geometry allows transmission and
superradiant amplification beyond this value. The difference comes from the region included in the wave problem and the boundary conditions imposed at
its endpoint.

\section{Discussion}
\label{sec:discussion}

For the parameter branch considered here, $k>0$ and $\Omega_i>\Omega_o$. Within the interval $0<\xi<1$, these conditions give $\nu_o>0$ and $\nu_i<0$.
Thus, a positive-norm wave incident from the outer-horizon side couples to a negative-norm outgoing channel at the neighbouring inner horizon.
Signed-flux conservation then allows the reflected flux to exceed the incident flux, producing superradiant amplification. The amplification band is
determined by the signs of the two horizon frequencies together with the transmission between the horizons. Its physical interpretation also depends on
the choice of incoming channel: the two-horizon problem and the transparent-end model describe different incident states and corresponding flux balances.

The radial potential also affects the scattering response. Local barriers can reduce transmission and produce resonant features through repeated
interactions within the radial region. The four lower-half-plane poles found in section~\ref{sec:poles} are damped resonances associated with the peaks
in the real-frequency response. Their positions show how the resonance damping changes with the background magnetic field and provide a spectral
interpretation of the observed amplification peaks. Confinement-induced superradiant instabilities in magnetic, mirror and anti-de Sitter settings
further illustrate the role of boundary conditions in determining the wave response
\cite{Konoplya:2008magnetic,Brito:2014magnetic,
Cardoso:2004nk,Cardoso:2004hs}. The geometric extension and outgoing spectral problem considered here follow the constructions of Zhou et
al.~\cite{Zhou:2026notes, Zhou:2026bridge}. The present analysis develops the bidirectional real-frequency response and relates its resonant features to
the local complex-frequency poles.

The ring singularity and the chronology-violating region are important features of the extended Kerr--BR geometry. For either independent analytic radial
solution, the energy on the positive-$r$ side of the ring is locally integrable, with
\begin{equation}
E_+(\epsilon)=O(\epsilon^{3/2}).
\end{equation}
This result characterizes the behaviour of the field on the chosen one-sided spacelike slice. A complete four-dimensional wave problem also requires a
specification of the allowed field behaviour at the ring and a suitable domain for the wave operator. In particular, the relation between the radial
solutions on the two sides of the ring is part of the global boundary-value problem. The distinction between radial extension, the choice of a wave
domain and well-posed evolution is central to the analysis of spacetimes with singularities or without global Cauchy
surfaces~\cite{Wald:1980nonglobally, IshibashiWald:2003dynamics,HorowitzMarolf:1995probes, KayRadzikowskiWald:1997cauchy}. The inner-horizon blue-shift
and backreaction also require a dynamical treatment beyond the stationary test-field flux analysis
\cite{PoissonIsrael:1990internal,Dafermos:2005interior}.
Accordingly, the gain $Z$ describes the flux amplification of the specified separated-mode problem.

Near-horizon methods provide another way to study wave dynamics in rotating black-hole geometries. The Kerr/CFT correspondence and near-extremal
scattering relate the response of the near-horizon region to a conformal description~\cite{Guica:2008mu,Bredberg:2009pv, Hartman:2009nz}. For Kerr--BR
spacetime, the corresponding near-horizon study derives symmetry and entropy properties~\cite{Siahaan:2025ngu}. The present work focuses on the
finite-spin, real-frequency scattering problem and its matching coefficients, which describe the response across the full two-horizon region.

The limit $B\to0$ requires separate consideration in the full connection problem. In this limit,
\begin{equation}
F(0)=B^2\alpha\longrightarrow0,
\end{equation}
and the finite-distance coordinate bridge approaches the Kerr asymptotic end. The local radial equation and angular eigenvalues provide checks against
the Kerr limit, while the global two-horizon connection matrix involves a different endpoint structure. Its relation to the usual Kerr
horizon-to-infinity scattering matrix therefore requires a separate limiting analysis. Similar extensions of the present study include extremal
backgrounds, axisymmetric modes and additional angular eigenvalue branches.

\section{Conclusions}
\label{sec:conclusions}

We have constructed the bidirectional real-frequency connection matrix for a neutral massless scalar field on the analytically extended Kerr--BR
geometry, using a specified two-horizon separated-mode problem. The matrix relates the incoming and outgoing amplitudes at the two horizons and satisfies
signed-flux conservation. The geometric extension and outgoing spectral setup follow the earlier analyses of the Kerr--BR spacetime
\cite{Zhou:2026notes,Zhou:2026bridge}. The main result
of this work is the real-frequency scattering response and its dependence on the retained geometry and endpoint conditions.

The central physical result is that superradiant amplification is determined by the signs of the two horizon frequencies together with the transmission
between the horizons. For positive-norm incidence from the left, amplification requires
\begin{equation}
\nu_o>0,\qquad \nu_i<0,\qquad T_L\ne0.
\end{equation}
More generally, the condition $\nu_o\nu_i<0$ identifies channels with opposite norm signs. The radial coefficient at the bridge can become negative while
transmission and amplification remain finite. This shows that a local potential barrier does not determine the response of the full radial region. The
cutoff near $BM_{\mathrm{ref}}=0.2427791435$ belongs to the transparent-end Robin model~\cite{Huang:2026transparent}, whereas the retained two-horizon
geometry supports amplification beyond this value.

We have also established the linear decrease of the gain towards both non-extremal horizon thresholds. The zero-shift Jost solutions provide an
independent calculation of the corresponding threshold slopes. For either independent analytic radial solution, the energy on the positive-$r$ side of
the ring is locally integrable, with
\begin{equation}
E_+(\epsilon)=O(\epsilon^{3/2}).
\end{equation}
Together, these results describe the threshold behaviour and local field properties of the two-horizon scattering problem.

Finally, the four simple lower-half-plane poles provide a local spectral interpretation of the real-frequency amplification peaks. As the background
magnetic field increases, these poles move farther into the lower half-plane, while the corresponding gain peaks become weaker. The real-frequency
response, the signed-flux relations and the nearby complex-frequency resonances therefore give complementary descriptions of wave scattering across the
Kerr--BR bridge. The resulting connection matrix characterizes the specified separated-mode continuation and provides a basis for further studies of the
global wave dynamics.

\section*{Code Availability Statement}

The Python codes used in this work are provided as electronic supplementary material accompanying this article. They allow the numerical calculations and figures presented in the manuscript to be independently reproduced.

\section*{Data Availability Statement}

All numerical data supporting the findings of this study can be generated using the accompanying Python codes. The package provides the necessary scripts
and instructions to reproduce the real-frequency scattering data, superradiant amplification curves, resonance frequencies.

\section*{Acknowledgments}
This work is supported by the National Natural Science Foundation of China under Grant Nos.~12373022 and U1731107, the Seventh Batch of High Level Innovative
Talents of Guizhou Province under Grant No.~GCC[2023]011, and the Special Funds for Discipline Construction and Postgraduate Education under Grant
No.~JX-2020-02.
\appendix

\section{Conformal separation and the physical current}
\label{app:separation}

Let $p_C=a\cos\theta$ and $\mathcal P=(a^2-p_C^2)\widetilde P$. The conformally related metric gives the following separated wave operator:
\begin{align}
\rho^2\left(\Box_{\widetilde g}
-\frac{R[\widetilde g]}6\right)
={}&\partial_r(Q\partial_r)
+\frac1{\sin\theta}\partial_\theta
\left(\widetilde P\sin\theta\,\partial_\theta\right)
\nonumber\\
&-\frac{[H\partial_t+a\partial_\varphi]^2}{Q}
+\frac{[a\sin^2\theta\,\partial_t+\partial_\varphi]^2}
{\widetilde P\sin^2\theta}
+\frac{Q''+\mathcal P_{,p_Cp_C}}6.
\label{eq:carter-operator}
\end{align}
The angular contribution from the last term is determined by
\begin{equation}
\frac{\mathcal P_{,p_Cp_C}}6
=-\frac13+\frac{\beta}{3}-2\beta\cos^2\theta.
\label{eq:carter-angular-term}
\end{equation}
Substituting the product mode for $\Psi=\Phi/\Omega$ into this operator separates the radial and angular equations, giving Eqs.\eqref{eq:angular} and
\eqref{eq:radial}. The conformal rescaling is essential for this separation. Without
it, the radial first-derivative coefficient contains
\begin{equation}
\partial_r\ln\Omega=
\frac{B^2[r-(\alpha r-d)\cos^2\theta]}{\Omega^2},
\end{equation}
which depends on both $r$ and $\theta$ for generic $B\ne0$. Consequently, a direct product ansatz for the unrescaled field $\Phi$ does not yield the
separated equations used in this work.

The physical scalar current is
\begin{equation}
j^\mu=
\frac{\Phi^*\nabla^\mu\Phi-
\Phi\nabla^\mu\Phi^*}{2\ii}.
\end{equation}
For the separated field, the relevant geometric and radial factors are
\begin{equation}
\sqrt{-g}=\Omega^{-4}\rho^2\sin\theta,\qquad
g^{rr}=\frac{\Omega^2Q}{\rho^2},
\end{equation}
and
\begin{equation}
\Phi^*\partial_r\Phi-\Phi\partial_r\Phi^*
=\Omega^2|S|^2(R^*R'-RR'^*).
\end{equation}
Because $\Omega$ is real, its radial derivatives cancel in the antisymmetric bilinear. The conformal factors from the volume element, inverse metric and
field bilinear also cancel in their product. After integration over the regular angular coordinate and the azimuthal period $2\pi C$, the physical radial
current therefore gives Eq.\eqref{eq:current} exactly. The same current is preserved across the coordinate bridge when the radial orientation is defined
by the common tortoise coordinate $s$.

The regular expression for the radial coefficient also removes apparent divergences in its original form. In Eq.\eqref{eq:potential}, the curvature and
derivative terms separately contain contributions proportional to $+2B^4\alpha^2r^2$ and $-2B^4\alpha^2r^2$. These terms cancel exactly, giving the
regular expression in Eq.\eqref{eq:regular-potential}. An independent representation is
\begin{equation}
V=\frac{N(r)}{H^4},\qquad
N=K^2H^2+Q(Q''/6-\lambda)H^2
-Q(Q+rQ')H+3r^2Q^2.
\end{equation}
After the cancellation of the highest-order terms, the numerator is a polynomial of degree eight:
\begin{equation}
N=\sum_{j=0}^8N_jr^j.
\label{eq:radial-numerator}
\end{equation}
For $M=1$, introduce the compact coordinate $r=\tan\chi$. The radial coefficient then takes the form
\begin{equation}
V(\chi)=
\frac{\displaystyle\sum_{j=0}^8
N_j\sin^j\chi\cos^{8-j}\chi}
{(\sin^2\chi+a^2\cos^2\chi)^4}.
\end{equation}
This expression and Eq.\eqref{eq:regular-potential} provide two independent forms of the radial coefficient for numerical integration. In the compact
coordinate, the expression can be evaluated directly at $\chi=\pi/2$, where $r$ diverges. This allows the radial equation to be evaluated at the
coordinate bridge without converting back to the singular coordinate $r$.

\section{Horizon initialization and amplitude extraction}\label{app:frobenius}
To relate the endpoint phases to the horizon coordinates, set $\phi_j=\varphi-\Omega_jt$ and $\delta\psi_j=\psi-\Omega_js$, which is smooth. With the
regular angles defined in section~\ref{sec:horizon-data}, the two radial signs give
\begin{align}
e^{-\ii\omega t+\ii k\varphi}e^{+\ii\nu_js}
&=e^{-\ii\nu_jw_-+\ii k\phi_j}
=e^{-\ii\nu_jw_-+\ii k\phi_j^u}e^{+\ii k\delta\psi_j},\\
e^{-\ii\omega t+\ii k\varphi}e^{-\ii\nu_js}
&=e^{-\ii\nu_jw_++\ii k\phi_j}
=e^{-\ii\nu_jw_++\ii k\phi_j^v}e^{-\ii k\delta\psi_j}.
\end{align}
The radial factors \(e^{+i\nu_j s}\) and \(e^{-i\nu_j s}\) correspond to \(u\)-type and \(v\)-type waves, respectively, for either sign of \(\nu_j\). At
fixed real
\((\omega,k)\), conjugating only the radial solution reverses the radial current and
propagation direction, while leaving \(\nu_j\) and the channel’s norm sign unchanged. This exchanges incoming and outgoing waves at the same endpoint.
Conjugating the full spacetime mode would also reverse \(\omega\) and \(k\). The negative-norm partner at the other endpoint is a separate input channel.

At either root, let $\delta=r-r_h$ and expand
\begin{equation}
Q=q_1\delta+q_2\delta^2+q_3\delta^3+\cdots,\qquad
K=k_0+k_1\delta+k_2\delta^2,
\end{equation}
with $q_j=Q^{(j)}(r_h)/j!$, $k_0=H_h\nu_h$, $k_1=2r_h\omega$, and $k_2=\omega$. Choose
\begin{equation}
R=\frac{|\delta|^\gamma}{\sqrt{H_h}}
\left(1+c_1\delta+c_2\delta^2+\cdots\right),\qquad
\gamma_o=-\frac{\ii k_{0,o}}{q_{1,o}},\qquad
\gamma_i=+\frac{\ii k_{0,i}}{q_{1,i}}.\label{eq:frobenius}
\end{equation}
At the neighbouring inner endpoint $\delta<0$. The modulus in $|\delta|^\gamma$ avoids an artificial exponential amplitude factor from a complex power of
a negative number. The signed $q_{1,i}$ already accounts for the endpoint orientation.

Multiplying the radial equation by $Q$ gives
\begin{equation}
Q^2R''+QQ'R'+[K^2+Q(Q''/6-\lambda)]R=0.
\end{equation}
The indicial equation is $q_1^2\gamma^2+k_0^2=0$. Define
\begin{align}
v_1&=2k_0k_1+q_1(q_2/3-\lambda),\\
v_2&=k_1^2+2k_0k_2+q_1q_3+q_2(q_2/3-\lambda).
\end{align}
The first two coefficients are
\begin{align}
c_1&=-\frac{q_1q_2\gamma(2\gamma+1)+v_1}{q_1^2(2\gamma+1)},\label{eq:frob-c1}\\
c_2&=-\frac{[q_1q_2(\gamma+1)(2\gamma+3)+v_1]c_1
+(q_2^2+2q_1q_3)\gamma(\gamma+1)+v_2}{4q_1^2(\gamma+1)}.\label{eq:frob-c2}
\end{align}
At order $N\ge1$, the recurrence denominator is $q_1^2N(N+2\gamma)$. It is nonzero for a non-extremal horizon and real nonzero frequency shift. At zero
shift the analytic branch remains, but the second independent solution becomes logarithmic; the incident/outgoing basis cannot then be inverted as two
distinct plane waves.

The initial state is $u=\sqrt H R$ and $u_s=(Q/H)\partial_r(\sqrt H R)$. The asymptotic normalization in Eq.\eqref{eq:frobenius} fixes $J[f_L]=-\noo$ and
$J[f_R]=\nii$.

For $W(f,g)=fg_s-f_sg$, let $\mathcal D=W(f_L,f_R)$. Eq.\eqref{eq:match} gives
\begin{equation}
T_L=\frac{2\ii\nu_o}{\mathcal D},\quad
T_R=\frac{2\ii\nu_i}{\mathcal D},\quad
R_L=-\frac{W(\overline f_L,f_R)}{\mathcal D},\quad
R_R=-\frac{W(f_L,\overline f_R)}{\mathcal D}.
\label{eq:wronskians}
\end{equation}
These formulas provide another way to extract the scattering amplitudes. Away from the horizon thresholds, a real-frequency zero of \(\mathcal D\) is a
scattering pole: it corresponds to a nonzero solution that is outgoing at both horizons. Such a zero is ruled out when \(\nu_o\nu_i>0\), because the
outgoing currents at the two ends would have opposite signs, contradicting current conservation. When
\(\nu_o\nu_i<0\), current conservation alone does not rule out such a pole. A finite
set of stable numerical results only checks the sampled frequencies and cannot establish that the whole interval is free of poles.

For arbitrary incoming amplitudes $a$ and outgoing amplitudes $b=\mathsf Sa$, the endpoint currents are
\begin{equation}
J_L=\noo(|a_L|^2-|b_L|^2),\qquad
J_R=\nii(|b_R|^2-|a_R|^2).
\end{equation}
Current conservation gives $a^\dagger\eta a=b^\dagger\eta b$ for every $a$, hence $\mathsf S^\dagger\eta\mathsf S=\eta$. This derivation requires real
frequency, nonzero endpoint shifts, the current-preserving continuation and a finite connection matrix.

\section{Exterior response and cut prescriptions}
\label{app:cut}

Consider a radial solution propagated between two finite matching sections. Its state vector $(u,v)$, with $v=u_s$, is related at the two sections by a
real fundamental matrix:
\begin{equation}
\binom{u_2}{v_2}=
\begin{pmatrix}
A_T&B_T\\
C_T&D_T
\end{pmatrix}
\binom{u_1}{v_1}.
\end{equation}
The response function $Y=u_s/u$ of this same solution therefore transforms as
\begin{equation}
Y_2=\frac{C_T+D_TY_1}{A_T+B_TY_1},
\qquad Y_j=\frac{v_j}{u_j}.
\end{equation}
Differentiating this relation along the radial coordinate gives the evolution equation in Eq.\eqref{eq:impedance}. When $u$ vanishes, $Y$ has a pole,
while the full state vector $(u,v)$ remains regular. Thus, propagation of the state vector preserves the same radial solution and its scattering
amplitudes at different matching sections, up to the chosen phase conventions.

A transparent boundary condition selects a particular response at the matching section. For $V>0$, the local outgoing-wave condition is $Y=+\ii\sqrt V$.
If this condition is imposed at each new cut $s_c$, the assigned response has derivative
\begin{equation}
\frac{\dd Y}{\dd s}
=\frac{\ii V_s}{2\sqrt V},
\end{equation}
whereas the response equation gives
\begin{equation}
Y_s=-V-Y^2=0
\end{equation}
for $Y=+\ii\sqrt V$ at that point. These relations generally differ when $V_s\ne0$. Consequently, imposing the same local transparent condition at
different cuts selects different exterior responses. The resulting scattering problem changes with the chosen cut.

The distinction is also important in a locally forbidden region, where $V<0$. A single decaying exponential has a real response $Y$ and carries zero
radial current. A combination of the two exponential solutions across a finite forbidden region, however, can carry nonzero current. The scattering
response therefore depends on the full solution across the region, including both exponential branches.

For a transparent cut with $q_c>0$, the local incoming and outgoing amplitudes are
\begin{equation}
A_{\rm in}=\frac12\left(u+\frac{\ii v}{q_c}\right),
\qquad
A_{\rm out}=\frac12\left(u-\frac{\ii v}{q_c}\right),
\qquad
J=q_c(|A_{\rm out}|^2-|A_{\rm in}|^2).
\end{equation}
At the fixed cut, define $J_0=\lim_{q_c\to0^+}J$ and $v_0=\lim_{q_c\to0^+}v$, where $v=u_s$ and both limits use the same smoothly normalized solution
family. At a nonresonant threshold with $v_0\ne0$, the gain follows the leading-order relation
\begin{equation}
Z_{\rm cut}=\frac{4J_0q_c}{|v_0|^2}+O(q_c^2).
\end{equation}
This relation describes the auxiliary channel at the transparent cut. The horizon-threshold relations in Eqs.\eqref{eq:outer-threshold} and
\eqref{eq:inner-threshold} describe the two horizon
channels, with their own endpoint modes and incident states. They therefore represent distinct threshold behaviours, each determined by its corresponding
scattering problem.

\section{One-sided energy near the ring}
\label{app:ring-energy}

We examine the local energy of the scalar field near the ring singularity from the positive-$r$ side. Set $x=\cos\theta$ and $\rho^2=r^2+a^2x^2$. In a
sufficiently small punctured neighbourhood with $r>0$, the conditions $Q>0$, $\widetilde P>0$, $\Omega^2>0$ and $\mathcal L>0$ ensure that the
constant-$t$ hypersurfaces are spacelike. Directly reducing the metric in Eq.\eqref{eq:metric} gives the lapse, the angular velocity of the normal
observers and their unit normal:
\begin{equation}
N=\frac{\rho\sqrt{Q\widetilde P}}{\Omega\sqrt{\mathcal L}},
\quad
\omega_{\rm Z}=\frac{a(\widetilde P H-Q)}{\mathcal L},
\quad
n=N^{-1}(\partial_t+\omega_{\rm Z}\partial_\varphi),
\quad
\dd\mathrm{vol}_h=
\frac{\rho}{\Omega^3}
\sqrt{\frac{\mathcal L}{Q\widetilde P}}\,
\dd r\,\dd x\,\dd\varphi.
\label{eq:ring-slice-data}
\end{equation}
Here $\dd\mathrm{vol}_h$ denotes the proper spatial volume element associated with the induced metric $h$ on the constant-$t$ hypersurface. The azimuthal
integration uses the period $2\pi C$.

The behaviour of the spatial geometry near the ring is controlled by $\mathcal L$. Its expansion is
\begin{align}
\mathcal L={}&2da^2r
+a^2(2-\alpha-B^2a^2)r^2
+a^4(1+\beta)x^2
+O\big((|r|+|x|)^3\big).
\label{eq:ring-L}
\end{align}
On the fixed rotating branch, $d>0$ and $1+\beta>0$. To make the dimensions explicit, introduce
\begin{equation}
\bar r=\frac rM,\qquad
\bar\rho=\frac{\rho}{M},\qquad
\bar a=\frac aM,\qquad
\bar{\mathcal L}=\frac{\mathcal L}{M^4}.
\label{eq:ring-dimensionless}
\end{equation}
In these variables, the local expansion becomes
\begin{align}
\bar{\mathcal L}={}&
2\frac dM\bar a^2\bar r
+\bar a^2(2-\alpha-B^2a^2)\bar r^2
+\bar a^4(1+\beta)x^2
+O\big((|\bar r|+|x|)^3\big).
\label{eq:ring-L-dimensionless}
\end{align}
Since $d>0$ and $1+\beta>0$, there are positive constants $c$ and $C$ such that, sufficiently close to the ring on the positive-$r$ side,
\begin{equation}
c(\bar r+x^2)
\leq\bar{\mathcal L}
\leq C(\bar r+x^2).
\label{eq:ring-L-bound}
\end{equation}
The constants depend on the fixed dimensionless background parameters. This bound is uniform throughout the neighbourhood, including the region
$\bar r\ll x^2$. Since $\bar\rho^2=\bar r^2+\bar a^2x^2$, it also gives
\begin{equation}
\bar{\mathcal L}=O(\bar\rho),
\qquad
\frac{\bar\rho^2}{\bar{\mathcal L}}=O(1).
\label{eq:ring-L-rho-bound}
\end{equation}
In dimensional variables, these relations imply $\mathcal L=O(M^3\rho)$ and $M^2\rho^2/\mathcal L=O(1)$. The normal-observer angular velocity
$\omega_{\rm Z}$ remains bounded as well.

We next use the local behaviour of the scalar field to estimate its energy density. The conformal factor satisfies $\Omega=1+O(\bar\rho^2)$. For either
analytic radial solution and regular angular branch, the field has the expansion
\begin{equation}
\Phi=e^{-\ii\omega t+\ii k\varphi}
\left[
R_0S_0+R_1S_0r+R_0S_1x+O(\rho^2)
\right].
\label{eq:ring-field-expansion}
\end{equation}
Here $R_0$ and $R_1$ are independent local radial data. The equatorially even dipole used in the numerical calculations has $S_1=0$, while the following
energy estimate also applies when $S_1\ne0$.

Treating the complex scalar as two real minimally coupled fields, we choose an overall normalization such that
\begin{equation}
T_{nn}=\frac12\left(
|n\Phi|^2+
h^{ab}\partial_a\Phi^*\partial_b\Phi
\right).
\label{eq:ring-stress-energy}
\end{equation}
The inverse spatial metric is diagonal in $(r,x,\varphi)$:
\begin{equation}
h^{rr}=\frac{\Omega^2Q}{\rho^2},\qquad
h^{xx}=\frac{\Omega^2\widetilde P(1-x^2)}{\rho^2},\qquad
h^{\varphi\varphi}
=\frac{\Omega^2\rho^2}{(1-x^2)\mathcal L}.
\label{eq:ring-inverse-metric}
\end{equation}
The radial and angular derivatives of the field expansion are bounded. The contribution from the normal derivative satisfies
\begin{equation}
|n\Phi|^2
=N^{-2}|\omega-k\omega_{\rm Z}|^2|\Phi|^2
\leq C\frac{\mathcal L}{\rho^2}.
\label{eq:ring-normal-derivative}
\end{equation}
Together with the bounded ratio $M^2\rho^2/\mathcal L$, these estimates give
\begin{equation}
T_{nn}\leq C\rho^{-2}.
\label{eq:ring-energy-density-bound}
\end{equation}

The spatial volume element in Eq.\eqref{eq:ring-slice-data} also obeys
\begin{equation}
\dd\mathrm{vol}_h
\leq CM^3\bar\rho^{3/2}\,
\dd\bar r\,\dd x\,\dd\varphi.
\label{eq:ring-volume-bound}
\end{equation}

To evaluate the energy near the ring, introduce the dimensionless transverse coordinates
\begin{equation}
\bar r=\bar\rho\cos\vartheta,\qquad
\bar a x=\bar\rho\sin\vartheta,\qquad
-\frac{\pi}{2}<\vartheta<\frac{\pi}{2}.
\label{eq:ring-transverse-coordinates}
\end{equation}
Their Jacobian is
\begin{equation}
\left|
\frac{\partial(\bar r,x)}
{\partial(\bar\rho,\vartheta)}
\right|
=\frac{\bar\rho}{|\bar a|}.
\label{eq:ring-transverse-jacobian}
\end{equation}
Combining the energy-density and volume-element bounds, and defining $\bar E_+=E_+/M$ and $\bar\epsilon=\epsilon/M$, the local energy on the positive-$r$
side satisfies
\begin{equation}
\bar E_+(\bar\epsilon)
\leq C\int_0^{\bar\epsilon}
\bar\rho^{1/2}\,\dd\bar\rho
=\frac{2C}{3}\bar\epsilon^{3/2}.
\label{eq:ring-energy-integrability}
\end{equation}
Equivalently, in dimensional variables,
\begin{equation}
E_+(\epsilon)
\leq CM\left(\frac{\epsilon}{M}\right)^{3/2}.
\label{eq:ring-energy-dimensional}
\end{equation}
Thus, although the local energy density can diverge as $\rho^{-2}$, its integral over the positive-$r$ region remains finite and vanishes at least as
fast as $\epsilon^{3/2}$ when the neighbourhood shrinks. The constants in these estimates depend on the mode and the fixed background, and the same
bounds need not hold uniformly as $a\to0$.

Since $R[g]=0$, minimally coupled and conformally coupled massless scalar fields obey the same wave equation. Their stress tensors, however, are
generally different, so the energy estimate above refers specifically to the minimally coupled stress tensor.

The geometry on the negative-$r$ side has a different causal structure near the ring. For small negative $r$ on the equatorial plane,
Eq.\eqref{eq:ring-L} gives
\begin{equation}
g_{\varphi\varphi}
=\frac{\mathcal L}{\Omega^2\rho^2}<0.
\label{eq:ring-azimuthal-metric}
\end{equation}
The azimuthal direction is then timelike, and the constant-$t$ hypersurface is no longer spacelike. The positive spatial metric and timelike unit normal
used above are therefore specific to the positive-$r$ side. Energy estimates on other spacelike hypersurfaces, together with the allowed field behaviour
at the ring, require a separate analysis. These questions determine how the local radial continuation is incorporated into a four-dimensional wave domain
and its time evolution.

\section{Real- and Complex-Frequency Analysis}
\label{app:denominator}

At the common matching section $r=0$, we introduce the dimensionless state columns
\begin{equation}
F_L=\binom{f_L}{M f_{L,s}},\qquad
F_R=\binom{f_R}{M f_{R,s}},\qquad
\mathsf A=(F_L,-F_R).
\label{eq:matching-matrix}
\end{equation}
The determinant satisfies $\det\mathsf A=-M\mathcal D$. In addition to the endpoint-normalized denominator $|M\mathcal D|$, we use the column-normalized
quantity
\begin{equation}
\widehat D=
\frac{|M\mathcal D|}
{\|F_L\|_2\|F_R\|_2},
\qquad
\widehat{\mathsf A}=
\left(
\frac{F_L}{\|F_L\|_2},
-\frac{F_R}{\|F_R\|_2}
\right).
\label{eq:normalized-denominator}
\end{equation}
For two unit columns, $0\leq\widehat D\leq1$, and the corresponding two-norm condition number is
\begin{equation}
\kappa_2(\widehat{\mathsf A})=
\frac{1+\sqrt{1-\widehat D^2}}{\widehat D}.
\label{eq:normalized-condition}
\end{equation}
This expression follows from the eigenvalues of the normalized Gram matrix. While $|M\mathcal D|$ retains the endpoint normalization, $\widehat D$
measures the relative alignment of the two matching columns and is invariant under their separate nonzero rescalings. The two quantities therefore
characterize complementary aspects of the matching problem. Their minima need not coincide.

Both the normalized denominator and the condition numbers depend on the choice of matching section. They serve as numerical diagnostics of the matching
problem rather than independent scattering observables. By contrast, the raw determinant is preserved under exact radial propagation, since the
first-order evolution matrix has zero trace and the associated transfer matrix has unit determinant.

We analyze the real-frequency response by locating and refining the minima of the endpoint-normalized and column-normalized denominators separately.
Their stability is assessed through independent radial integrations and changes in numerical resolution. This provides a consistent characterization of
the real-frequency matching structure.

For complex frequencies, we continue the angular eigenvalue branch associated with the real-frequency dipole mode through the generalized Galerkin
eigenproblem
\begin{equation}
[A_0+\omega A_1+\omega^2 A_2]c
=\lambda(\omega)\,G c.
\label{eq:complex-angular}
\end{equation}
Here $G$ is the positive real mass matrix, and the operator is generally non-Hermitian for complex $\omega$. The relevant eigenvalue is selected by
continuous tracking from the real-frequency branch. Using this angular eigenvalue, we analytically continue the outgoing Frobenius solutions and their
Wronskian in the complex-frequency plane.

The radial conjugation relations in Eq.~\eqref{eq:wronskians} are restricted to real frequencies. Consequently, the complex-frequency problem is
formulated directly in terms of the analytically continued matching determinant $\mathcal D(\omega)$. Its zeros identify the candidate scattering poles,
whereas $\widehat D$ and the real-frequency gain are not analytic pole conditions.

We focus on the complex-frequency neighbourhood associated with the resolved real-frequency response peaks. The resulting roots are examined through
their local stability and the behaviour of the analytically continued determinant. This connects the real-frequency scattering response to the nearby
complex-frequency pole structure.

\bibliographystyle{references}
\bibliography{refs_verified}

@article{Bertotti:1959uniform,
  author  = {Bertotti, B.},
  title   = {Uniform electromagnetic field in the theory of general relativity},
  journal = {Physical Review},
  volume  = {116},
  pages   = {1331--1333},
  year    = {1959},
  doi     = {10.1103/PhysRev.116.1331}
}

@article{Robinson:1959maxwell,
  author  = {Robinson, I.},
  title   = {A solution of the {Maxwell--Einstein} equations},
  journal = {Bulletin de l'Acad{\'e}mie Polonaise des Sciences, S{\'e}rie des Sciences Math{\'e}matiques, Astronomiques et Physiques},
  volume  = {7},
  pages   = {351--352},
  year    = {1959}
}

@article{Podolsky:2025tle,
    author = "Podolsky, Jiri and Ovcharenko, Hryhorii",
    title = "{Kerr Black Hole in a Uniform Bertotti-Robinson Magnetic Field: An Exact Solution}",
    eprint = "2507.05199",
    archivePrefix = "arXiv",
    primaryClass = "gr-qc",
    doi = "10.1103/rfgv-ybz5",
    journal = "Phys. Rev. Lett.",
    volume = "135",
    number = "18",
    pages = "181401",
    year = "2025"
}

@article{Ovcharenko:2026genuine,
    author = {Ovcharenko, Hryhorii and Podolsk{\'y}, Ji{\v r}{\'i}},
    title = {{Kerr-Newman and Genuine Kerr Black Holes in the Bertotti-Robinson Magnetic Field}},
    journal = {Phys. Rev. Lett.},
    volume = {137},
    pages = {201403},
    year = {2026},
    doi = {10.1103/skq4-mn78},
    eprint = {2608.21672},
    archivePrefix = {arXiv},
    primaryClass = {gr-qc}
}

@article{Carter:1968separable,
  author  = {Carter, Brandon},
  title   = {Hamilton--Jacobi and {Schr{\"o}dinger} separable solutions of {Einstein}'s equations},
  journal = {Communications in Mathematical Physics},
  volume  = {10},
  pages   = {280--310},
  year    = {1968},
  doi     = {10.1007/BF03399503}
}

@article{Teukolsky:1973perturbations,
  author  = {Teukolsky, Saul A.},
  title   = {Perturbations of a rotating black hole. {I}. Fundamental equations for gravitational, electromagnetic, and neutrino-field perturbations},
  journal = {The Astrophysical Journal},
  volume  = {185},
  pages   = {635--647},
  year    = {1973},
  doi     = {10.1086/152444}
}

@article{FrolovKrtousKubiznak:2017hidden,
  author  = {Frolov, Valeri P. and Krtou{\v{s}}, Pavel and Kubiz{\v{n}}{\'a}k, David},
  title   = {Black holes, hidden symmetries, and complete integrability},
  journal = {Living Reviews in Relativity},
  volume  = {20},
  pages   = {6},
  year    = {2017},
  doi     = {10.1007/s41114-017-0009-9},
  eprint  = {1705.05482},
  archivePrefix = {arXiv}
}

@article{Gray:2025lwy,
    author = "Gray, Finnian and Kubiznak, David and Ovcharenko, Hryhorii and Podolsky, Jiri",
    title = "{Hidden symmetries and separability structures of Ovcharenko-Podolsk{\'y} and conformal-to-Carter spacetimes}",
    eprint = "2511.21538",
    archivePrefix = "arXiv",
    primaryClass = "gr-qc",
    doi = "10.1103/8832-htpg",
    journal = "Phys. Rev. D",
    volume = "113",
    number = "4",
    pages = "044050",
    year = "2026"
}

@article{Starobinskii:1974amplification,
  author  = {Starobinskii, A. A. and Churilov, S. M.},
  title   = {Amplification of electromagnetic and gravitational waves scattered by a rotating ``black hole''},
  journal = {Soviet Physics JETP},
  volume  = {38},
  pages   = {1--5},
  year    = {1974}
}

@article{Unruh:1974secondquantization,
  author  = {Unruh, William G.},
  title   = {Second quantization in the {Kerr} metric},
  journal = {Physical Review D},
  volume  = {10},
  pages   = {3194--3205},
  year    = {1974},
  doi     = {10.1103/PhysRevD.10.3194}
}

@article{TeukolskyPress:1974interaction,
  author  = {Teukolsky, Saul A. and Press, William H.},
  title   = {Perturbations of a rotating black hole. {III}. Interaction of the hole with gravitational and electromagnetic radiation},
  journal = {The Astrophysical Journal},
  volume  = {193},
  pages   = {443--461},
  year    = {1974},
  doi     = {10.1086/153180}
}

@book{FuttermanHandlerMatzner:1988scattering,
  author    = {Futterman, J. A. H. and Handler, F. A. and Matzner, R. A.},
  title     = {Scattering from Black Holes},
  publisher = {Cambridge University Press},
  year      = {1988},
  isbn      = {9780511735615},
  doi       = {10.1017/CBO9780511735615}
}

@article{Glampedakis:2001cx,
    author = "Glampedakis, Kostas and Andersson, Nils",
    title = "{Scattering of scalar waves by rotating black holes}",
    eprint = "gr-qc/0102100",
    archivePrefix = "arXiv",
    doi = "10.1088/0264-9381/18/10/309",
    journal = "Class. Quant. Grav.",
    volume = "18",
    pages = "1939--1966",
    year = "2001"
}

@article{Unruh:1976absorption,
  author  = {Unruh, William G.},
  title   = {Absorption cross section of small black holes},
  journal = {Physical Review D},
  volume  = {14},
  pages   = {3251--3259},
  year    = {1976},
  doi     = {10.1103/PhysRevD.14.3251}
}

@article{Page:1976emission,
  author  = {Page, Don N.},
  title   = {Particle emission rates from a black hole. {II}. Massless particles from a rotating hole},
  journal = {Physical Review D},
  volume  = {14},
  pages   = {3260--3273},
  year    = {1976},
  doi     = {10.1103/PhysRevD.14.3260}
}

@article{Macedo:2013afa,
    author = "Macedo, Caio F. B. and Leite, Lu{\'\i}z C. S. and Oliveira, Ednilton S. and Dolan, Sam R. and Crispino, Lu{\'\i}s C. B.",
    title = "{Absorption of planar massless scalar waves by Kerr black holes}",
    eprint = "1308.0018",
    archivePrefix = "arXiv",
    primaryClass = "gr-qc",
    doi = "10.1103/PhysRevD.88.064033",
    journal = "Phys. Rev. D",
    volume = "88",
    number = "6",
    pages = "064033",
    year = "2013"
}

@article{Brito:2015oca,
    author = "Brito, Richard and Cardoso, Vitor and Pani, Paolo",
    title = "{Superradiance -- the 2020 Edition}",
    eprint = "1501.06570",
    archivePrefix = "arXiv",
    primaryClass = "gr-qc",
    doi = "10.1007/978-3-030-46622-0",
    journal = "Lect. Notes Phys.",
    volume = "971",
    year = "2020"
}

@article{Wald:1974np,
    author = "Wald, Robert M.",
    title = "{Black hole in a uniform magnetic field}",
    doi = "10.1103/PhysRevD.10.1680",
    journal = "Phys. Rev. D",
    volume = "10",
    pages = "1680--1685",
    year = "1974"
}

@article{Ernst:1976mzr,
    author = "Ernst, Frederick J.",
    title = "{Black holes in a magnetic universe}",
    doi = "10.1063/1.522781",
    journal = "J. Math. Phys.",
    volume = "17",
    number = "1",
    pages = "54--56",
    year = "1976"
}

@article{Ernst:1976bsr,
    author = "Ernst, Frederick J. and Wild, Walter J.",
    title = "{Kerr black holes in a magnetic universe}",
    doi = "10.1063/1.522875",
    journal = "J. Math. Phys.",
    volume = "17",
    number = "2",
    pages = "182--184",
    year = "1976"
}

@article{Aliev:1989wx,
    author = "Aliev, A. N. and Galtsov, D. V.",
    title = "{Magnetized Black Holes}",
    doi = "10.1070/PU1989v032n01ABEH002677",
    journal = "Sov. Phys. Usp.",
    volume = "32",
    pages = "75",
    year = "1989"
}

@article{Konoplya:2008magnetic,
  author  = {Konoplya, R. A.},
  title   = {Superradiant instability for black holes immersed in a magnetic field},
  journal = {Physics Letters B},
  volume  = {666},
  pages   = {283--287},
  year    = {2008},
  doi     = {10.1016/j.physletb.2008.07.079},
  eprint  = {0801.0846},
  archivePrefix = {arXiv},
  primaryClass = {hep-th},
  note = {Erratum: Phys. Lett. B 670 (2009) 459, doi: 10.1016/j.physletb.2008.11.059.}
}

@article{Brito:2014magnetic,
  author  = {Brito, Richard and Cardoso, Vitor and Pani, Paolo},
  title   = {Superradiant instability of black holes immersed in a magnetic field},
  journal = {Physical Review D},
  volume  = {89},
  pages   = {104045},
  year    = {2014},
  doi     = {10.1103/PhysRevD.89.104045},
  eprint  = {1405.2098},
  archivePrefix = {arXiv},
  primaryClass = {gr-qc}
}

@article{Gibbons:2013yq,
    author = "Gibbons, G. W. and Mujtaba, A. H. and Pope, C. N.",
    title = "{Ergoregions in Magnetised Black Hole Spacetimes}",
    eprint = "1301.3927",
    archivePrefix = "arXiv",
    primaryClass = "gr-qc",
    reportNumber = "DAMTP-2013-5, MIFPA-13-02",
    doi = "10.1088/0264-9381/30/12/125008",
    journal = "Class. Quant. Grav.",
    volume = "30",
    number = "12",
    pages = "125008",
    year = "2013"
}

@article{Cardoso:2004nk,
    author = "Cardoso, Vitor and Dias, Oscar J. C. and Lemos, Jose P. S. and Yoshida, Shijun",
    title = "{The Black hole bomb and superradiant instabilities}",
    eprint = "hep-th/0404096",
    archivePrefix = "arXiv",
    doi = "10.1103/PhysRevD.70.044039",
    journal = "Phys. Rev. D",
    volume = "70",
    pages = "044039",
    year = "2004",
    note = "[Erratum: Phys.Rev.D 70, 049903 (2004)]"
}

@article{Cardoso:2004hs,
    author = "Cardoso, Vitor and Dias, Oscar J. C.",
    title = "{Small Kerr-anti-de Sitter black holes are unstable}",
    eprint = "hep-th/0405006",
    archivePrefix = "arXiv",
    doi = "10.1103/PhysRevD.70.084011",
    journal = "Phys. Rev. D",
    volume = "70",
    pages = "084011",
    year = "2004"
}

@article{Wang:2025kbrshadow,
    author = "Wang, Xinyu and Hou, Yehui and Wan, Xi and Guo, Minyong and Chen, Bin",
    title = "{Geodesics and Shadows in the Kerr--Bertotti--Robinson Black Hole Spacetime}",
    eprint = "2507.22494",
    archivePrefix = "arXiv",
    primaryClass = "gr-qc",
    doi = "10.1088/1475-7516/2026/02/050",
    journal = "JCAP",
    volume = "02",
    pages = "050",
    year = "2026"
}

@article{Wan:2026kbrphotonrings,
    author = "Wan, Xi and Zhang, Zhenyu and Wei, Fang-Stars and Hou, Yehui and Chen, Bin",
    title = "{Critical behavior of photon rings in Kerr--Bertotti--Robinson spacetime}",
    journal = "Phys. Rev. D",
    volume = "114",
    pages = "044005",
    year = "2026",
    doi = "10.1103/fyky-bkbg",
    eprint = "2603.25049",
    archivePrefix = "arXiv",
    primaryClass = "gr-qc"
}

@article{Mustafa:2026gly,
    author = "Mustafa, G. and Donmez, Orhan and Gogoi, Dhruba Jyoti and Ghosh, Sushant G. and Hussain, Ibrar and Yuan, Chengxun",
    title = "{Dynamics, Ringdown, and Accretion-Driven Multiple Quasi-Periodic Oscillations of Kerr-Bertotti-Robinson Black Holes}",
    eprint = "2602.08911",
    archivePrefix = "arXiv",
    primaryClass = "gr-qc",
    month = "2",
    year = "2026"
}

@article{Xu:2026ags,
    author = "Xu, Hengyu and Chen, Haowei and Zhang, Shao-Jun",
    title = "{Horizon-Evanescent Scalar Clouds from Coupled Rotation and Magnetic Fields around Black Holes}",
    eprint = "2606.27958",
    archivePrefix = "arXiv",
    primaryClass = "gr-qc",
    doi = "10.1007/JHEP09(2026)283",
    journal = "JHEP",
    volume = "09",
    pages = "283",
    month = "6",
    year = "2026"
}

@article{Ma:2026ima,
    author = "Ma, Liang and Lu, H.",
    title = "{New Rotating Black Hole in Electromagnetic Fields: Cosmological Horizon without Cosmological Constant}",
    eprint = "2606.23782",
    archivePrefix = "arXiv",
    primaryClass = "gr-qc",
    month = "6",
    year = "2026"
}

@article{Ma:2026otg,
    author = "Ma, Liang and Lu, H.",
    title = "{Demagnetizing KBR and New Ricci-flat Rotating Metric}",
    eprint = "2605.13954",
    archivePrefix = "arXiv",
    primaryClass = "gr-qc",
    month = "5",
    year = "2026"
}

@article{Ovcharenko:2026global,
    author = {Ovcharenko, Hryhorii},
    title = {{Investigation of global structure of Kerr-Newman-Bertotti-Robinson spacetime}},
    journal = {Phys. Rev. D},
    volume = {114},
    pages = {104045},
    year = {2026},
    doi = {10.1103/z77w-tld7},
    eprint = {2608.28265},
    archivePrefix = {arXiv},
    primaryClass = {gr-qc}
}

@article{Zhou:2026notes,
  author        = {Zhou, Yu-Sen and Wu, Liang-Bi and Ji, Ming-Fei and Fu, Wen-Tao and Cao, Li-Ming and Cai, Rong-Gen},
  title         = {Notes on {Kerr--Bertotti--Robinson} Spacetime},
  year          = {2026},
  eprint        = {2608.19136},
  archivePrefix = {arXiv},
  primaryClass  = {gr-qc},
  url           = {https://arxiv.org/abs/2608.19136},
  note          = {Short report; arXiv:2609.07602 is identified by its authors as the companion detailed study.}
}

@article{Zhou:2026bridge,
  author        = {Zhou, Yu-Sen and Wu, Liang-Bi and Ji, Ming-Fei and Fu, Wen-Tao and Cao, Li-Ming and Cai, Rong-Gen},
  title         = {Global structure and stability of {Kerr--Bertotti--Robinson} spacetime},
  year          = {2026},
  eprint        = {2609.07602},
  archivePrefix = {arXiv},
  primaryClass  = {gr-qc},
  url           = {https://arxiv.org/abs/2609.07602}
}

@article{Huang:2026transparent,
  author        = {Huang, Hai and Sun, Xudong and Chen, Juhua},
  title         = {Massless scalar scattering by {Kerr--Bertotti--Robinson} black holes: transparent-end channels and superradiance},
  year          = {2026},
  eprint        = {2607.16068},
  archivePrefix = {arXiv},
  primaryClass  = {gr-qc},
  url           = {https://arxiv.org/abs/2607.16068}
}

@article{Wald:1980nonglobally,
  author  = {Wald, Robert M.},
  title   = {Dynamics in nonglobally hyperbolic, static space-times},
  journal = {Journal of Mathematical Physics},
  volume  = {21},
  pages   = {2802--2805},
  year    = {1980},
  doi     = {10.1063/1.524403}
}

@article{IshibashiWald:2003dynamics,
  author  = {Ishibashi, Akihiro and Wald, Robert M.},
  title   = {Dynamics in non-globally-hyperbolic static spacetimes: {II}. General analysis of prescriptions for dynamics},
  journal = {Classical and Quantum Gravity},
  volume  = {20},
  pages   = {3815--3826},
  year    = {2003},
  doi     = {10.1088/0264-9381/20/16/318},
  eprint  = {gr-qc/0305012},
  archivePrefix = {arXiv}
}

@article{HorowitzMarolf:1995probes,
  author  = {Horowitz, Gary T. and Marolf, Donald},
  title   = {Quantum probes of spacetime singularities},
  journal = {Physical Review D},
  volume  = {52},
  pages   = {5670--5675},
  year    = {1995},
  doi     = {10.1103/PhysRevD.52.5670},
  eprint  = {gr-qc/9504028},
  archivePrefix = {arXiv}
}

@article{Berti:2005gp,
    author = "Berti, Emanuele and Cardoso, Vitor and Casals, Marc",
    title = "{Eigenvalues and eigenfunctions of spin-weighted spheroidal harmonics in four and higher dimensions}",
    eprint = "gr-qc/0511111",
    archivePrefix = "arXiv",
    doi = "10.1103/PhysRevD.73.024013",
    journal = "Phys. Rev. D",
    volume = "73",
    pages = "024013",
    year = "2006",
    note = "[Erratum: Phys.Rev.D 73, 109902 (2006)]"
}

@article{Carter:1968global,
  author  = {Carter, Brandon},
  title   = {Global structure of the {Kerr} family of gravitational fields},
  journal = {Physical Review},
  volume  = {174},
  pages   = {1559--1571},
  year    = {1968},
  doi     = {10.1103/PhysRev.174.1559}
}

@book{Chandrasekhar:1983mathematical,
  author    = {Chandrasekhar, S.},
  title     = {The Mathematical Theory of Black Holes},
  publisher = {Clarendon Press},
  address   = {Oxford},
  year      = {1983},
  isbn      = {0198512910}
}

@article{KayRadzikowskiWald:1997cauchy,
  author  = {Kay, Bernard S. and Radzikowski, Marek J. and Wald, Robert M.},
  title   = {Quantum field theory on spacetimes with a compactly generated {Cauchy} horizon},
  journal = {Communications in Mathematical Physics},
  volume  = {183},
  pages   = {533--556},
  year    = {1997},
  doi     = {10.1007/s002200050042},
  eprint  = {gr-qc/9603012},
  archivePrefix = {arXiv}
}

@article{PoissonIsrael:1990internal,
  author  = {Poisson, Eric and Israel, Werner},
  title   = {Internal structure of black holes},
  journal = {Physical Review D},
  volume  = {41},
  pages   = {1796--1809},
  year    = {1990},
  doi     = {10.1103/PhysRevD.41.1796}
}

@article{Dafermos:2005interior,
  author  = {Dafermos, Mihalis},
  title   = {The interior of charged black holes and the problem of uniqueness in general relativity},
  journal = {Communications on Pure and Applied Mathematics},
  volume  = {58},
  pages   = {445--504},
  year    = {2005},
  doi     = {10.1002/cpa.20071},
  eprint  = {gr-qc/0307013},
  archivePrefix = {arXiv}
}

@article{Guica:2008mu,
    author = "Guica, Monica and Hartman, Thomas and Song, Wei and Strominger, Andrew",
    title = "{The Kerr/CFT Correspondence}",
    eprint = "0809.4266",
    archivePrefix = "arXiv",
    primaryClass = "hep-th",
    doi = "10.1103/PhysRevD.80.124008",
    journal = "Phys. Rev. D",
    volume = "80",
    pages = "124008",
    year = "2009"
}

@article{Bredberg:2009pv,
    author = "Bredberg, Irene and Hartman, Thomas and Song, Wei and Strominger, Andrew",
    title = "{Black Hole Superradiance From Kerr/CFT}",
    eprint = "0907.3477",
    archivePrefix = "arXiv",
    primaryClass = "hep-th",
    doi = "10.1007/JHEP04(2010)019",
    journal = "JHEP",
    volume = "04",
    pages = "019",
    year = "2010"
}

@article{Hartman:2009nz,
    author = "Hartman, Thomas and Song, Wei and Strominger, Andrew",
    title = "{Holographic Derivation of Kerr-Newman Scattering Amplitudes for General Charge and Spin}",
    eprint = "0908.3909",
    archivePrefix = "arXiv",
    primaryClass = "hep-th",
    doi = "10.1007/JHEP03(2010)118",
    journal = "JHEP",
    volume = "03",
    pages = "118",
    year = "2010"
}

@article{Siahaan:2025ngu,
    author = "Siahaan, Haryanto M.",
    title = "{Kerr{\textendash}Bertotti{\textendash}Robinson spacetime and the Kerr/CFT correspondence}",
    eprint = "2512.12533",
    archivePrefix = "arXiv",
    primaryClass = "gr-qc",
    doi = "10.1016/j.nuclphysb.2026.117514",
    journal = "Nucl. Phys. B",
    volume = "1028",
    pages = "117514",
    year = "2026"
}
\end{document}